\documentstyle[times,epsfig,graphics,amssymb,astrobib]{mn2e}

\newcommand{\be}{\begin{equation}}
\newcommand{\e}{\end{equation}}
\newcommand{\bear}{\begin{eqnarray}}
\newcommand{\ear}{\end{eqnarray}}

\newcommand{\f}{\frac}
\newcommand{\de}{{\rm d}}

\title[Constraining the reionization history with QSO absorption spectra]
{Constraining the reionization history with QSO absorption spectra}
\author[Gallerani, Choudhury \& Ferrara]
{S. Gallerani\thanks{E-mail: galleran@sissa.it}, 
T. Roy Choudhury\thanks{E-mail: chou@sissa.it},
and
A. Ferrara\thanks{E-mail: ferrara@sissa.it}\\
SISSA/ISAS, via Beirut 2-4, 34014 Trieste, Italy.}

\begin{document}

\maketitle

 \begin{abstract}
   We use a semi-analytical approach to simulate absorption spectra of
   QSOs at high redshifts with the aim of constraining the cosmic
   reionization history.  We consider two physically motivated and
   detailed reionization histories: (i) an Early Reionization Model
   (ERM) in which the intergalactic medium is reionized by PopIII
   stars at $z\approx 14$, and (ii) a more standard Late Reionization
   Model (LRM) in which overlapping, induced by QSOs and normal
   galaxies, occurs at $z\approx 6$.  From the analysis of current
   Ly$\alpha$ forest data at $z < 6$, we conclude that it is impossible
   to disentangle the two scenarios, which fit equally well the
   observed Gunn-Peterson optical depth, flux probability distribution
   function and dark gap width distribution.  At $z>6$, however, clear
   differences start to emerge which are best quantified by the dark
   gap and peak width distributions.  We find that 35 (zero)
   per cent of the lines of sight within $5.7< z <6.3$ show dark gaps
   widths $>50$\AA \ in the rest frame of the QSO if reionization is
   not (is) complete at $z \gtrsim 6$.  Similarly, the ERM predicts
   peaks of width $\sim 1$\AA \ in 40 per cent of the lines of sight
   in the redshift range $6.0-6.6$; in the same range, LRM predicts no
   peaks of width $>0.8$\AA. We conclude that the dark gap and peak
   width statistics represent superb probes of cosmic reionization if
   about ten QSOs can be found at $z > 6$. We finally discuss strengths
   and limitations of our method.
\end{abstract}


\section{Introduction}

After the recombination epoch at $z\sim 1100$, the universe remained
almost neutral until the first generation of luminous sources (stars,
accreting black holes, etc) were formed. The photons from these sources
ionized the surrounding neutral medium and once these individual
ionized regions started overlapping, the global ionization and thermal
state of the intergalactic gas changed drastically. This is known as
the reionization of the universe, which has been an important subject
of research over the last few years, especially because of its strong
impact on the formation and evolution of the first cosmic structures
(for a comprehensive review on the subject of reionization and first
cosmic structures, see \citeNP{cf05a}).  Calculations based on the
hierarchical structure formation in cold dark matter (CDM) models,
predict that reionization should naturally occur somewhere between $z
\sim 6-15$ \cite{co00,gnedin00}.

Recent observational progresses, however, seem to indicate that the
reionization history at high redshifts might have been a complicated
process.  To start with, WMAP observations of the
temperature-polarization cross correlation at large scales for the CMB
suggest a Thomson optical depth as high as $\tau_e\sim 0.17$, which
implies that reionization might have occurred at redshifts as high as
$z\sim 15$ \cite{ksb++03,svp++03}. On the other hand, the spectroscopy
of the Ly$\alpha$ forest for QSOs at $z > 6$ discovered by the Sloan
Digital Sky Survey (SDSS; \citeNP{fnl+01,fss++03,f05}) seems to indicate
that the ionization state of the intergalactic medium (IGM) might be
very different along different lines of sight.  For example, the
analyses of the spectrum of the most distant known quasar (SDSS
J1148+5251) show some residual flux both in the Ly$\alpha$ and
Ly$\beta$ troughs, which when combined with Ly$\gamma$ region
\cite{fo05}, imply that this flux is consistent with pure
transmission.  The presence of unabsorbed regions in the spectrum
corresponds to a highly ionized IGM along that particular line of
sight.  However, \citeN{bfw++01} detected a complete Gunn-Peterson
trough in the spectrum of SDSS J1030+0524 ($z=6.28$), where no
transmitted flux is detected over a large region (300 \AA) immediately
blueward of the Ly$\alpha$ emission line.  This result has been shown
to be consistent with a hydrogen neutral fraction $f_{\rm HI} \gtrsim
10^{-3}$ (\citeNP{fnswbpr02}, hereafter F02).

There have been other different approaches to investigate the neutral
hydrogen fraction. \citeN{wlc05} and \citeN{wl04} have estimated the
sizes of the ionized regions around 7 QSOs at $z > 6$ (which included
the above cited QSO).  Following that they considered the neutral gas
surrounding the QSO as a function of different parameters: the
Str\"omgren sphere size $R_S$, the quasar's production rate of
ionizing photons $\dot{N}_{\rm phot}$, the clumping factor of the gas
$C$ and the age of the QSO $t_{\rm age}$. According to their
arguments, the small sizes of the HII regions ($\sim 10$ physical Mpc)
imply that the typical neutral hydrogen fraction of the IGM beyond
$z\sim 6$ is in the range 0.1 - 1.  However, this approach is weighted
down by several uncertainties.  For example, one of the uncertainties
is the quasar's production rate of ionizing photons $\dot{N}_{\rm
  phot}$ as it depends on the shape of the spectral template used.
Moreover it is implicitly assumed in the modelling of clumping factor
that the formation of quasars and galaxies were simultaneous. This in
turn implies that quasars ionize only low density regions and hence
the clumping factor which regulates the evolution of the HII regions
is low. If, instead, stars appears much earlier than QSOs, the quasars
have to ionize high density regions, which means that one should use a
higher value of clumping factor in the calculations \cite{yl05}.
   
\citeN{mh04} have used a different approach based on the damping wings
of the neutral hydrogen. Using density and velocity fields obtained by
hydrodynamical simulation, they computed the Ly$\alpha$ absorption as
a function of wavelength. In this case the neutral hydrogen fraction,
$\dot{N}_{\rm phot}$ and $R_S$ are treated as free parameters,
constrained by matching the optical depth observed in the QSO SDSS
J1030+0524. Also in this case they find a neutral hydrogen fraction
larger than 10 per cent, i.e. the IGM is significantly more neutral at
$z\sim 6$ than the lower limit directly obtainable from the GP trough
of the QSO spectrum ($10^{-3}$).  However this result is based only on
one quasar. Moreover the observational constraints on the optical
depth are very uncertain and can introduce errors in the estimates of
$f_{\rm HI}$.

The fact that we find transmission along some lines of sight while the
medium seems quite neutral along others possibly has been interpreted
that the IGM ionization properties are different along different lines
of sight at $z \gtrsim 6$, thus suggesting that we might be observing
the end of the reionization process.

Finally, the evolution of the luminosity function of Ly$\alpha$
emitters \cite{maro04,fzh05,hcen05} suggests that the neutral fraction
of hydrogen at $z=6.5$ should be greater than 50 per cent
\cite{maro05}.

From a theoretical point of view several efforts have been made in the last 
few years in order to reconcile WMAP and SDSS measurements 
(e.g. \citeNP{cfo03,hh03,gnedin04}). 
In particular, Choudhury \& Ferrara (2005, hereafter CF05) 
\nocite{cf05} have showed 
that a self-consistent model for reionization, which agrees with 
various sets of observations over
a wide redshift range, predicts a highly ionized universe at $z
\approx 6$.  According to the model, the rise in the GP optical depth
towards $z=6$ is achieved by assuming a drop of the photoionization
rate caused by the disappearance of first generation metal free
(hereafter PopIII) stars.  To explore this idea, it becomes important
to verify whether the difference in the GP trough along different
lines of sight at $z > 6$ can be explained by assuming a highly
ionized universe. In other words we pose the question: is it possible
that the IGM is overall in a highly ionized state and the differences
observed in QSO spectra arise simply due to the cosmic variance in
density fluctuations?

Since at redshifts larger than 5, the transmission in the spectra is
extremely low, the mean transmitted flux being $F_{\rm mean} \lesssim
0.182$ (\citeNP{songaila04}, hereafter S04), it is not possible to
study the properties of the Ly$\alpha$ forest through the usual Voigt
profile analysis.  Alternative quantities have been used to
characterize the spectra and to compare them with models, such as the
probability distribution function (PDF) of the flux
(F02; \citeNP{sc02}, hereafter SC02) and more importantly, the
distribution of the dark gap widths.  It turns out that the length of
dark gaps at high redshifts can be quite large ($\sim 80$ Mpc
comoving, which corresponds to $\sim 30$\AA \ in the rest frame
wavelength at redshift 6), and hence the modelling requires a large
sample of very long lines of sight (say, $\gtrsim 100$ Mpc).  Such
requirement is beyond the reach of current numerical simulations and
can only be fulfilled through semi-analytical calculations.

In this paper we use semi-analytical techniques to produce artificial
spectra of the Ly$\alpha$ forest in the redshift range $5.0<z<6.6$. We
have considered different reionization scenarios, namely, (i) an early
reionization model (CF05), characterized by an highly ionized IGM for
$z \lesssim14$, and (ii) a late reionization model, in which the
overlapping of ionized regions is completed only by $z\sim 6$.  The
main goal of this paper is to identify a statistics for the Ly$\alpha$
forest at $z > 6$ which can be used as an effective tool for
discriminating between the two scenarios above.

The outline of this paper is as follows. Section 2 presents the
formalism to simulate the Ly$\alpha$ flux from the IGM density and
peculiar velocity fields. In addition, we also describe the two
reionization scenarios in detail. Section 3 contains the comparison of
our models with observational data at $z < 6$.  We introduce various
statistical tools for the Ly$\alpha$ forest and use our models to make
testable predictions for $z \gtrsim 6$. Finally we summarize our
conclusions in Section 4.

\section{Simulating the QSO absorption spectra}

The Ly$\alpha$ forest, observed in the absorption spectra of distant
quasars, is the result of photon absorption through Ly$\alpha$
transition of neutral hydrogen gas in the intergalactic medium along
the line of sight (LOS).  It is believed that the Ly$\alpha$ forest
arises from low-amplitude fluctuations in the underlying baryonic
density field \cite{bbc92}. In this work, we shall use the formalism 
developed by \citeN{bd97}, \citeN{csp01} and \citeN{vmht02} to simulate 
random LOS realizations of the Ly$\alpha$ forest as observed in the quasar
absorption spectra.

\subsection{Neutral hydrogen density and the transmitted flux}

Let us first summarize the steps for calculating the 
neutral hydrogen density field, optical depth and the transmitted 
flux for the Ly$\alpha$ forest:

\subsubsection{Linear density and velocity fields for baryons}
In the linear regime, both the density and velocity fields for baryons
are gaussian and are completely determined by the corresponding power
spectra and correlation functions.  The linear density power spectrum
for baryons in one dimension is given by
\begin{equation}
P_b^{(1)}(k,z)=\frac{D^2(z)}{2\pi}
\int_k^{\infty} dq ~ q ~ W_b^2(q,z) ~ P_{\rm DM}^{(3)}(q)
\end{equation}
where $P_{\rm DM}^{(3)}(k)$ is the dark matter power spectrum in three
dimensions at redshift $z = 0$, $D(z)$ is the growth factor of linear
DM density fluctuations [normalized so that $D(z=0) = 1$] and
$W_b(k,z)$ is the low-pass filter which suppresses baryonic
fluctuations at small scales. It is well known that the exact form of
$W_b(k,z)$ depends on the ionization and thermal history of the
universe, and one should, in principle, couple the evolution of the
baryonic fluctuations to the reionization history. Since this is
computationally quite complex, one is led to use various approximate
forms for the filter function. It turns out that if the temperature of
the IGM is smoothly increasing with redshift and does not undergo any
abrupt change (which is the case for our models), the filter function
can be assumed to be of the form 
\begin{equation} 
W_b(k,z) = \f{1}{1+x_b^2(z) k^2}
\end{equation} 
where $x_b(z)$ denotes the comoving scale below which fluctuations
are suppressed:
\begin{equation}\label{jsc}
x_b(z)=\frac{1}{H_0}
\left[\frac{2\gamma k_B T_m(z)}{3\mu m_p \Omega_m (1+z)}\right]^{1/2}.
\end{equation}
The constant $\mu$ is the mean molecular weight of the IGM, given by 
$\mu=2/(4-3Y)$, where $Y=0.24$ is the helium mass fraction (this relation 
assumes that the IGM consist mostly of fully ionized hydrogen and singly 
ionized helium). In equation (\ref{jsc}) $k_B$ is the Boltzmann constant, 
$m_p$ is the hydrogen mass, $\Omega_m$ is the density parameter for 
non-relativistic matter, $T_m$ is the mass-averaged temperature and $\gamma$ 
is the ratio of specific heats.
Note that we are using the mass-averaged temperature $T_m$ 
which is considerably
higher than the conventionally used temperature at the mean gas density
$T_0$. This is 
motivated by the fact that baryonic density fluctuations calculated using the 
mass-averaged temperature are in a 
much better agreement with the results of numerical simulations. The evolution 
of $T_m$ with $z$ has been computed as described in CF05, in Section 5.

The CDM power spectrum in three dimensions is taken to be 
$P_{\rm DM}^{(3)}(k)=A_{\rm DM}~k^n~T_{\rm DM}^2(q)$, where $n$ is the 
spectral index and $T_{\rm DM}(q)$ is the CDM transfer function \cite{bbks86}:
\begin{eqnarray}
&&T_{\rm DM}(q)=\frac{\ln(1+2.34q)}{2.34q}\times\nonumber\\
&&\Big[1+3.89q+(16.1q)^2+(5.46q)^3+(6.71q)^4\Big]^{-1/4},
\end{eqnarray}
with $q \equiv k/(h~{\rm Mpc}^{-1}) \Gamma^{-1}$. The shape parameter $\Gamma$ 
depends on the Hubble parameter, $\Omega_m$ and $\Omega_b$ \cite{sugiyama95}:
\begin{equation}
\Gamma=\Omega_mh \exp\left[-\Omega_b
\left(1+\f{\sqrt{2 h}}{\Omega_m}\right)\right].
\end{equation}
The normalization parameter $A_{\rm DM}$ is fixed through the value of 
$\sigma_8$ (the rms density fluctuations in spheres of radius 8 $h^{-1}$Mpc).

The computation of the linear 
baryonic peculiar velocity field $v_{\rm pec}(x,z)$ is 
similar to that for the density field.
We use the linear power spectrum for the 
velocity field in one dimension given by
\begin{equation}
P_v^{(1)}(k,z)= \left[\f{\dot{D}(z)}{1+z}\right]^2
k^2
\frac{1}{2\pi}\int_k^{\infty}\frac{\de q}{q^3}W_b^2(q,z)
P_{\rm DM}^{(3)}(q)
\end{equation}
In addition, we do take into account the fact that the 
velocity field is correlated with the density field 
\begin{equation}
P_{bv}^{(1)}(k,z) = \f{\dot{D}(z)}{1+z}  k
\frac{1}{2\pi}\int_k^{\infty}\frac{\de q}{q}W_b^2(q,z)
P_{\rm DM}^{(3)}(q).
\end{equation}
Given the above relations and using the properties
of gaussian random fields, one can generate the 
density and peculiar velocity fields for baryons in the linear regime.
These
are discussed in details in 
\citeN{bd97}, \citeN{csp01} and \citeN{vmht02}.

\subsubsection{Quasi-linear density field for baryons} 

The above analysis is done in the framework of linear perturbation theory, 
while to study the properties of the IGM one has to take into account 
nonlinearities in the density distribution.
To generate the mildly non-linear regime of the IGM local density we use a 
lognormal model, introduced by \citeN{cj91} for the nonlinear matter 
distribution in the universe and widely adopted later 
\cite{bi93,bd97,cps01,csp01,vmht02}
for the case of IGM. 
The lognormal distribution of the baryonic density field is given by
\begin{equation}
n_b(x,z)=
n_0(z) \exp \left[\delta_b(x,z)-\frac{\langle\delta_b(x,z)\rangle^2}{2}\right]
\label{ln}
\end{equation}
where $\delta_b$ is the linear baryonic density contrast and $n_0(z)$ is the 
mean IGM density which is related to the baryonic density 
parameter $\Omega_b$ and the critical density $\rho_c$ through the relation
\begin{equation}
n_0(z)=\frac{\Omega_b\rho_c}{\mu m_p}(1+z)^3.
\end{equation}

We should mention here that the lognormal model is found to under-predict the 
number of high density regions compared to what is found in numerical
simulations (CF05), and hence is inadequate
for describing the highly non-linear densities. 
However, since these high density regions are quite rare, we expect them to 
have little effect on the statistics of Ly$\alpha$ forest performed in this 
paper. Thus, as far as this work is concerned, the lognormal approximation 
should work reasonably well for studying the properties of Ly$\alpha$ 
forest. The accuracy of the lognormal
approximation has been validated by \citeN{csp01} and CF05; the model is able 
to match various sets 
of observations simultaneously over a wide range of redshifts. 
For the purpose of this paper, we have also carried out some 
additional comparisons of the lognormal model with HydroPM
simulations \cite{gh98} and found that, as far as the flux statistics 
used in this paper (the PDF and the distribution of dark gap widths) 
are concerned, the model gives quite 
good agreement with simulations. The details of such comparisons
are given in Appendix \ref{lognormal}.

\subsubsection{Neutral hydrogen distribution}

The low density gas which gives rise to the Ly$\alpha$ forest is approximately 
in local equilibrium between photoionization and recombination, expressed by 
the relation
\begin{equation}\label{photoeq}
\alpha[T(x,z)] n_p(x,z) n_e(x,z)= \Gamma_{\rm HI}(x,z) n_{\rm HI}(x,z)
\end{equation} 
where $\alpha(T)$ is the radiative recombination rate, $n_e$ and $n_p$ are the 
number density of electrons and protons respectively and $\Gamma_{\rm HI}$ is 
the photoionization rate of neutral hydrogen. 
The characteristic low density of the IGM allow us to neglect the collisional 
ionizations.

It is useful to define the neutral hydrogen fraction $f_{\rm HI}$: 
\begin{equation}
f_{\rm HI}(x,z) \equiv \frac{n_{\rm HI}(x,z)}{n_H(x,z)}
= 1.08 \frac{n_{\rm HI}(x,z)}{n_b(x,z)}.
\end{equation}
where the factor 1.08 arises because of the presence of helium.
To solve the ionization equilibrium equation (\ref{photoeq}) 
in an exact manner we 
need to know the precise ionization state of helium (which affects the
number density of electrons $n_e$). However
we can obtain the neutral fraction in the two extreme cases which are discussed
next.
Usually, the ionization conditions in the Ly$\alpha$ forest at $3.5< z < 5.5$ 
are similar to those of HII regions with $f_{\rm HI} \lesssim 10^{-4}$; 
furthermore
in such epochs, helium is mostly
in a singly-ionized state. Thus with the approximation
$f_{\rm HI} \ll 1$, equation (\ref{photoeq}) gives
\begin{equation}
f_{\rm HI}(x,z) \approx 1.08 \frac{\alpha[T(x,z)]}{\Gamma_{\rm HI}(x,z)} 
n_H(x,z)= 
\frac{\alpha[T(x,z)]}{\Gamma_{\rm HI}(x,z)} n_b(x,z). 
\end{equation}
However, at higher redshifts (say, $z > 5.5$), one has to consider the 
possibility that the IGM is not completely ionized and there remain regions
with a high neutral fraction.
Such regions are opaque to ionizing radiation, and hence the 
effective photoionization rate in such regions can be taken to be zero; 
it follows that $f_{\rm HI} \approx 1$.

The recombination coefficient at temperature T is given by \cite{rms++97}
\begin{equation}
\alpha[T(x,z)]=4.2 \times 10^{-13}\left[\frac{T(x,z)}{10^4K}\right]^{-0.7}
{\rm cm}^3 {\rm s}^{-1}
\end{equation} 
For quasi-linear IGM, where non-linear effects like shock-heating can be 
neglected, the temperature $T(x,z)$ can be related to the baryonic density 
through a power-law relation \cite{hg97,stres00}
\begin{equation}
T(x,z)=T_0(z) \left[\frac{n_b(x,z)}{n_0(z)}\right]^{\gamma-1}
\end{equation}
where $T_0(z)$ is the temperature of the IGM at the mean density. 

The slope $\gamma$ of the equation of state depends on the reionization 
history of the universe \cite{tlept98,hg97}. The value of $\gamma$ and its 
evolution are still quite uncertain and hence in this work we will consider 
it as a free parameter and ignore its redshift evolution.

\subsubsection{Optical depth and transmitted flux for the 
Ly$\alpha$ forest}
\label{voigt}
The transmitted flux $F$ due to Ly$\alpha$ absorption in the IGM 
is computed from the usual relation $F=e^{-\tau_{{\rm Ly} \alpha}}$, 
where $\tau_{{\rm Ly} \alpha}$ is the  Ly$\alpha$ absorption optical depth. 
The value of the optical depth
at a redshift $z_0$ is given by
\begin{eqnarray}
\tau_{{\rm Ly} \alpha}(z_0)
&=&\f{c I_{\alpha}} {\sqrt{\pi}} \int~{\rm d} x(z)
\f{n_{\rm HI}(x(z),z)}{b(x(z),z)(1+z)} \nonumber \\ 
&\times& V\left[\alpha,\f{v_H(z_0) - v_H(z) - v_{\rm pec}(x(z),z)}
{b(x(z),z)}\right],
\end{eqnarray}
where $v_H(z_0) - v_H(z) = c (z_0 - z)/(1 + z_0)$ 
denotes the differential Hubble velocity
between two points along the LOS.
The quantity $I_{\alpha} = 4.45 \times 10^{-18}$ cm$^2$, 
\begin{equation}
b(x,z) = \left[\f{2 k_B T(x,z)}{m_p}\right]^{1/2}
\end{equation} 
is the Doppler parameter, $V$ is the Voigt function and $\alpha$ measures 
the natural line width of  Ly$\alpha$ transition. The 
quantity $x(z)$ denotes the comoving distance to a point 
along the LOS at a redshift $z$:
\begin{equation}
x(z) = \int_0^z \de z' \f{c}{H(z')}
\end{equation}

In this work, each LOS is discretized in a number of pixels $N_{\rm pix}$. 
At each pixel, we calculate the neutral hydrogen density $n_{\rm HI}$ and 
the peculiar velocity $v_{\rm pec}$ through the procedure 
discussed above. 
Then the optical depth at a pixel $i$ is given by
\begin{equation}
\tau_{{\rm Ly} \alpha}(i)=c I_{\alpha} \f{\Delta x}{1+z}
\sum^{N_{\rm pix}}_{j=1} n_{\rm HI}(j) \Phi_{\alpha}[v_H(i)-v(j)]
\label{lyalphatau}
\end{equation}
where $\Delta x$ is the comoving pixel size, 
$v(i)=v_H(i)+v_{\rm pec}(i)$ is the total velocity in the pixel $i$ and
$\Phi_{\alpha}$ is the Voigt profile for Ly$\alpha$ transition.
For low column density 
regions, the natural broadening is 
not that important, and the Voigt function reduces to 
a simple Gaussian
\begin{equation}
\Phi_{\alpha}[v_H(i)-v(j)]=\frac{1}{\sqrt{\pi}~b(j)} 
\exp\left[-\left(\f{v_H(i)-v(j)}{b(j)}\right)^2\right].
\end{equation}
However, one should keep in mind that while dealing with highly neutral
regions (which is relevant for the late reionization scenario), the 
Gaussian approximation for the line profile is not valid, and
one has to use the appropriate form for the Voigt profile. In regions away
from the center, the profile is given by the Lorentzian form:
\begin{equation}
\Phi_{\alpha}[v_H(i)-v(j)]=\frac{R_{\alpha}}{\pi [(v_H(i)-v(j))^2
+ R_{\alpha}^2]}
\end{equation}
where $R_{\alpha} \equiv \Lambda_{\alpha} \lambda_{\alpha}/4 \pi$ with 
$\Lambda_{\alpha}$ being the decay constant for the Ly$\alpha$ resonance
and $\lambda_{\alpha}$ is the wavelength of the Ly$\alpha$ line.

Similar expressions follow for Ly$\beta$ absorption lines too. In 
particular equation (\ref{lyalphatau}) is replaced by
\begin{equation}
\tau_{{\rm Ly} \beta}(i)=c I_{\beta} \f{\Delta x}{1+z}
\sum^{N_{\rm pix}}_{j=1} n_{\rm HI}(j) \Phi_{\beta}[v_H(i)-v(j)]
\label{lybetatau}
\end{equation}
where $I_{\beta} = (f_{{\rm Ly}\beta}\lambda_{{\rm Ly}\beta})/
(f_{{\rm Ly}\alpha}\lambda_{{\rm Ly}\alpha}) I_{\alpha}$, with 
$(f_{{\rm Ly}\beta}\lambda_{{\rm Ly}\beta})/
(f_{{\rm Ly}\alpha}\lambda_{{\rm Ly}\alpha})=0.16$ being the 
ratio of the product between the oscillator strength and the 
resonant scattering wavelength for Ly$\beta$ and Ly$\alpha$;
$\Phi_{\beta}$ is the Voigt profile for Ly$\beta$ transition.
For low column density systems, $\Phi_{\beta}$ has the Gaussian
form and it is essentially the same as $\Phi_{\alpha}$. Hence, for such
systems, we have $\tau_{{\rm Ly} \beta}(i) = 0.16 \tau_{{\rm Ly} \alpha}(i)$.
However, the situation is different for highly neutral regions, where
$\Phi_{\beta}$ depends on the decay constant for the resonance line and
thus is different from $\Phi_{\alpha}$. In particular, for regions away
from the center, the profile is of the Lorentzian form:
\begin{equation}
\Phi_{\beta}[v_H(i)-v(j)]=\frac{R_{\beta}}{\pi [(v_H(i)-v(j))^2
+ R_{\beta}^2]}
\end{equation}
where $R_{\beta} \equiv \Lambda_{\beta} \lambda_{\beta}/4 \pi$. 

\subsubsection{Simulation of observational and instrumental effects}

In this work, we compute various statistical quantities 
related to the Ly$\alpha$ 
forest using our simulated spectra, such as (i) the evolution of the 
Gunn-Peterson optical depth ($\tau_{GP}$), (ii) 
the Probability Distribution Function (PDF) and 
(iii) the Dark Gap Width Distribution (DGWD), which can 
then be compared with observational results.
To make sure that the simulated spectra contain the same artifacts
as the observed ones, we take into account the 
broadening of lines due to instrumental profile, 
the pixel size and noise.
In this regard, 
we first convolve each simulated spectra with a Gaussian with a 
Full Width at Half Maximum (FWHM) corresponding to the resolution 
of the instrument used for observations. We than re-sample each line 
to varying pixel size. 
We finally add noise to the simulated Ly$\alpha$ forest 
spectra  corresponding to the observed data in a manner that
the flux $F$ in each pixel is replaced by 
$F \to F + \sigma_{\rm noise} G(1)$, where $\sigma_{\rm noise} = 0.02$
and $G(1)$ is a Gaussian random deviate with zero mean and unit
variance.
We have also studied the effect of varying the FWHM, the pixel size
and $\sigma_{\rm noise}$ on different statistical quantities and shall comment
on them wherever appropriate.

\subsection{Reionization models}

As it is clear from the above discussion, the simulation of 
the Ly$\alpha$ forest spectra requires the knowledge of a few  
free parameters. For the background cosmological 
model and the CDM power spectrum, we use the best-fit 
values given by WMAP experiment\footnote{Throughout this paper we will assume 
a flat universe with total matter, vacuum, and baryonic densities in units of 
the critical density of $\Omega_m=0.27$, $\Omega_{\Lambda}=0.73$, and 
$\Omega_bh^2=0.024$, respectively, and a Hubble constant of 
$H_0=100 h$~km~s$^{-1}$Mpc$^{-1}$, with $h=0.72$. The parameters defining the 
linear dark matter power spectrum are $\sigma_{8}=0.9$, $n=0.99$, 
$\de n/\de \ln k=0$.}. This leaves us
with the parameters related to the IGM, which are
$T_m(z), T_0(z), \Gamma_{\rm HI}(x,z)$ and $\gamma(z)$. 
One approach could be to treat them as free parameters
and try to constrain them by comparison with observations. However,
the evolution of all the above parameters 
depends on the detailed ionization and thermal history of the 
IGM and can be quite complex -- hence 
constraining the parameters over a wide redshift range is not straightforward.
The other approach is to use a self-consistent model for
thermal and ionization history of the universe and calculate
the globally-averaged values of the above quantities. 

In this paper, we take the second approach and 
use the semi-analytical model of CF05
to obtain the globally-averaged values of $T_m, T_0, \Gamma_{\rm HI}$ at
different redshifts. The model implements most of
the relevant physics governing the thermal and ionization
history of the IGM, such as the
inhomogeneous IGM density distribution, three different classes of
ionizing photon sources (massive PopIII stars, PopII stars and QSOs),
and radiative feedback inhibiting star formation in low-mass
galaxies. The main advantage of the model is that its parameters 
can be constrained
quite well by comparing its predictions 
with various observational data, namely, 
the redshift evolution of Lyman-limit absorption systems \cite{smih94},
Gunn-Peterson \cite{songaila04} and electron scattering optical depths
\cite{ksb++03}, temperature of the IGM \cite{stle99} and cosmic star formation 
history \cite{nchos04}.

According to the above model, the redshift of reionization 
is identified with the onset of the 
post-overlap stage \cite{gnedin00}, which is defined as the 
epoch where the volume filling factor of ionized hydrogen 
in low-density regions (with overdensities less than a few tens)
reaches unity ($Q_{\rm HII}=1$). Following this, the 
ionized regions start propagating into the neutral high density regions,
which is manifested as the evolution in the specific number of Lyman-limit
systems.
In what follows we will consider two different reionization scenarios:

\vspace{0.2cm}
\centerline{\it Early Reionization Model (ERM)} 

\noindent This model, which refers to the fiducial model described in CF05, 
is characterized by an highly ionized IGM 
at redshifts $z \lesssim 14$. In this scenario, an early population
of massive metal-free (PopIII) stars ionize hydrogen at high redshifts,
thus producing the high electron scattering optical depth 
observed by WMAP. The PopIII stars start disappearing 
at $z_{\rm trans} \approx 10$; 
at lower redshifts, 
PopII stars and QSOs contribute to the ionizing background 
with a large number of photons with energies above 13.6 eV which 
are able to maintain the ionized state of hydrogen. 
In this model we assume the photoionization rate to be spatially
homogeneous and equal to the globally averaged value, i.e., 
$\Gamma_{\rm HI}(x,z) \equiv \Gamma_{\rm HI}(z)$

In the context of the present work, where we are concerned 
with state of the IGM at $5.5 \lesssim z \lesssim 6.5$, the ERM
corresponds to a highly ionized IGM at these redshifts. While
it is true that this model can explain a large number of observational
constraints, it is in contradiction with the analyses 
predicting that the IGM could be in a highly neutral state
at $z \gtrsim 6$ \cite{wlc05,wl04,mh04}. Hence it becomes 
necessary to consider an alternative
model for reionization where the IGM is predominantly neutral 
at $z \gtrsim 6$.

\vspace{0.2cm}
\centerline{\it Late Reionization Model (LRM)}  

\noindent The main motivation to consider this model is to verify 
whether the Ly$\alpha$ forest can still be used to determine the 
ionization state of the IGM at $z \gtrsim 6$.
In this model, the hydrogen distribution in the 
low-density IGM is characterized by two distinct phases
at $z \gtrsim 6$, namely an ionized (HII) phase with a 
volume filling factor $Q_{\rm HII}$ and a
neutral (HI) phase with a 
volume filling factor $1 - Q_{\rm HII}$, with the
evolution of $Q_{\rm HII}$ and other physical parameters, 
[$T_m(z), T_0(z), \Gamma_{\rm HI}(z)$]
being calculated self-consistently using the model of CF05.
To achieve this two-phased state, we consider an ionizing background different 
from ERM one. In fact, 
the main (and only) difference between ERM and LRM is that the 
PopIII stars do not play an efficient role 
for reionization in LRM and as a consequence the 
IGM remains neutral up to redshift 
6, until the contribution of PopII stars to the UV background 
starts becoming substantial. 
In passing, we should mention that in the ERM the electron scattering 
optical depth is $\tau_e=0.17$, in perfect agreement 
with the high value measured by WMAP, 
while in the LRM it is $\tau_e=0.06$, value which is lower than the 
2$\sigma$ limit allowed by WMAP.

\vspace{0.2cm}
\centerline{\it Model comparison}

\begin{figure*}
\centerline{
\psfig{figure=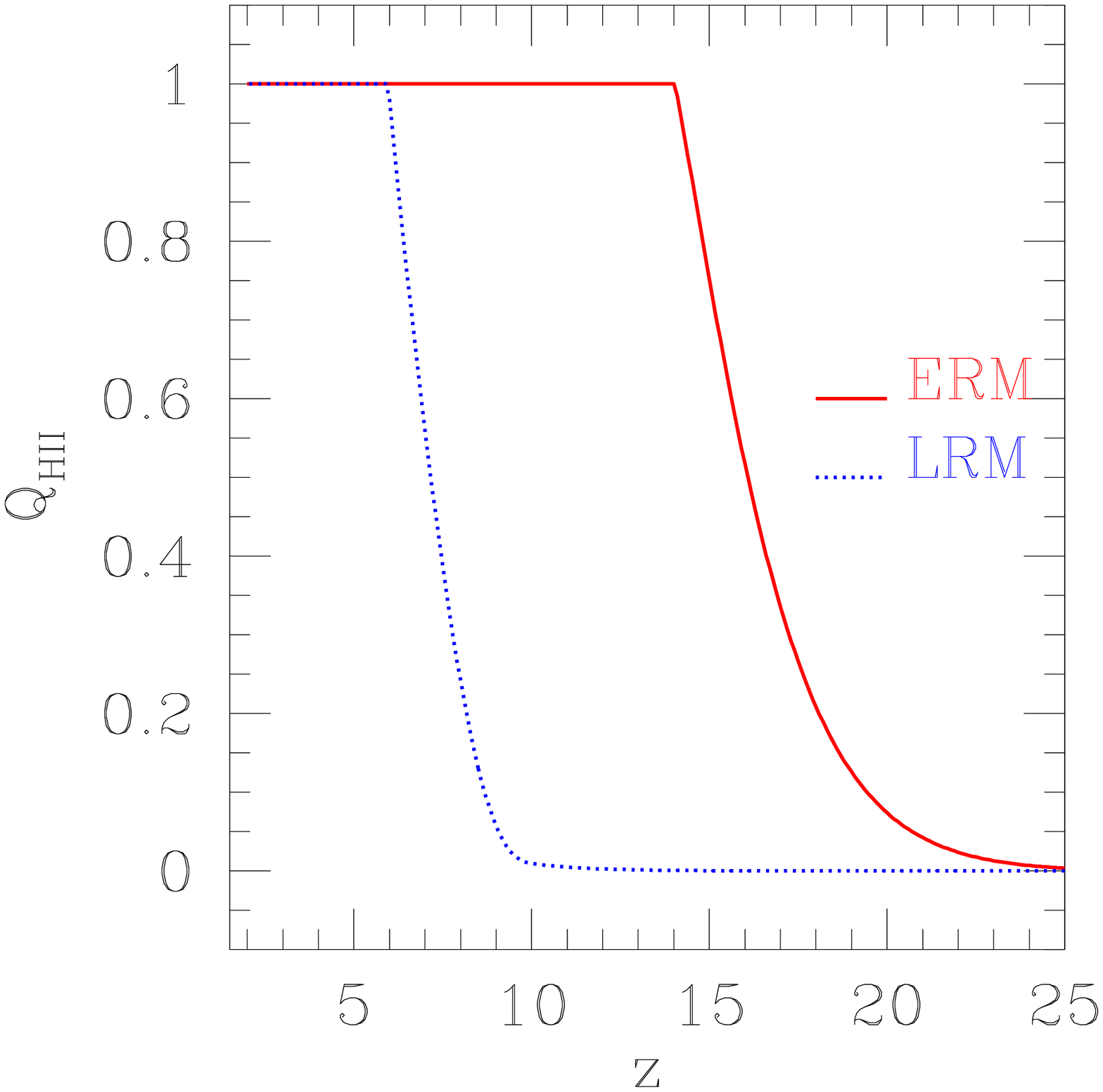,width=6.5cm,angle=0}
$\!\!\!\!\!\!\!\!\!\!\!$
\psfig{figure=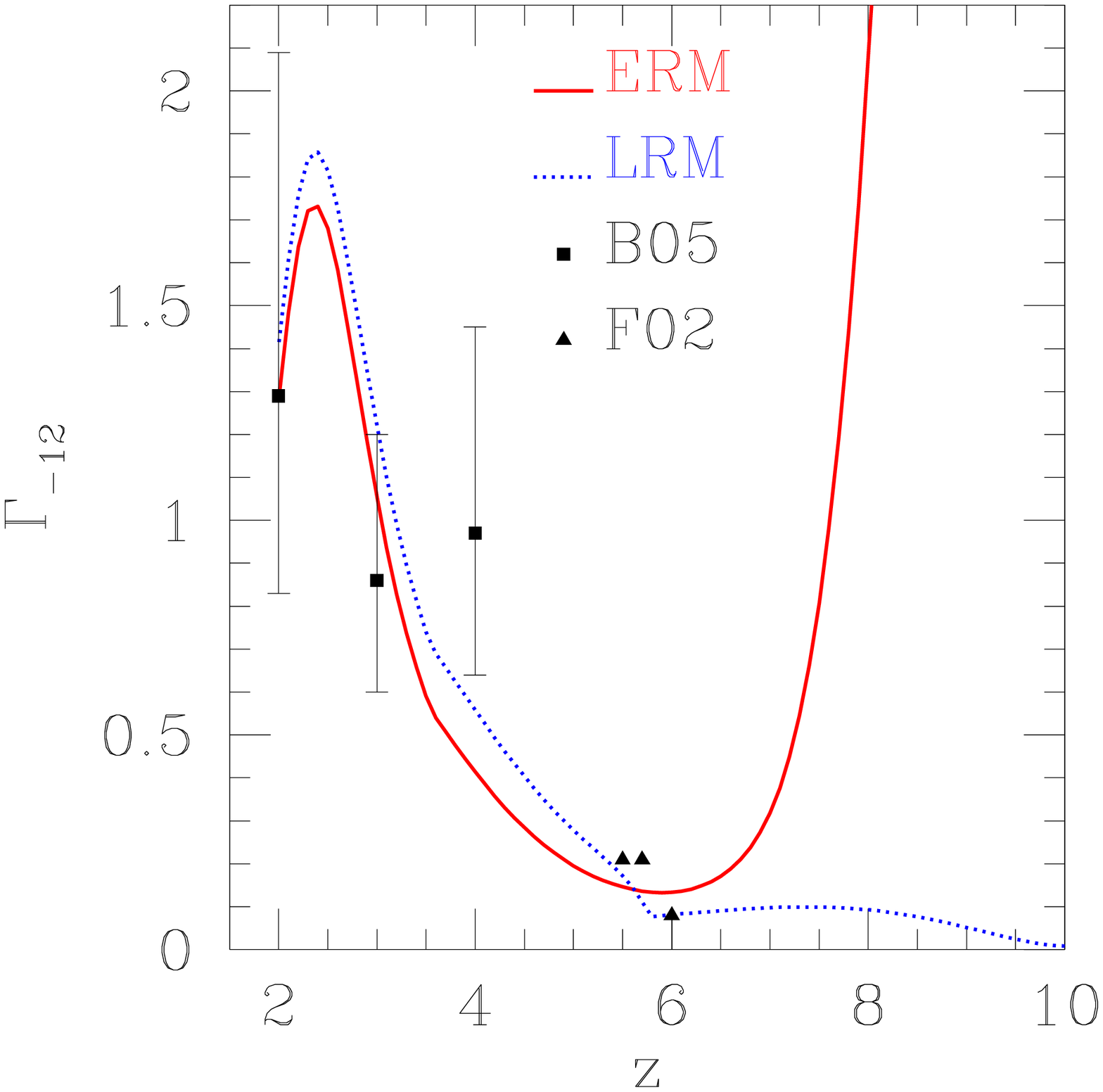,width=6.5cm,angle=0}
$\!\!\!\!\!\!\!\!\!\!\!$
\psfig{figure=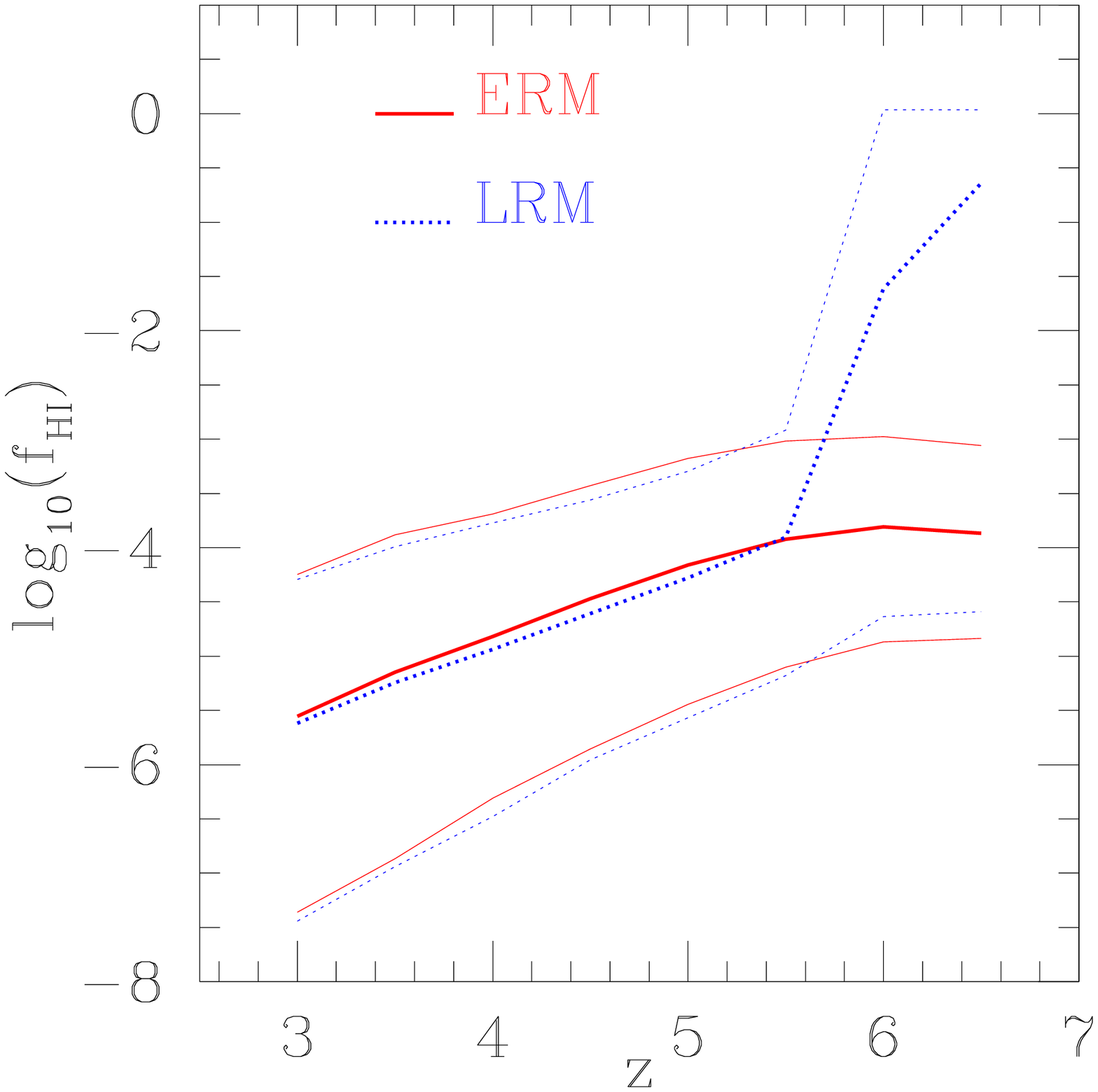,width=6.5cm,angle=0}
}
\caption{Evolution of the volume filling 
factor of ionized regions (left), the globally volume-averaged 
photoionization rate in units of $10^{-12}\rm{s^{-1}}$, 
$\Gamma_{-12}=\Gamma_{\rm HI}/10^{-12}{\rm s}^{-1}$(middle) and the neutral 
hydrogen 
fraction (right)
for the early (ERM) and late (LRM) reionization models. 
The points in the middle panel show results obtained by Bolton et al.(2005)
 (B05, filled squares) and Fan et al.(2002) (F02, filled triangles) using 
hydrodynamical and N-body simulations, respectively.
In the right panel, thick lines represent average results over 10 LOS, 
while the thin lines denote the upper and lower neutral hydrogen fraction 
extremes in each redshift interval.}
\label{Q_z}
\end{figure*}

\noindent To compare the global properties of the two reionization models, 
we plot the evolution of the volume filling factor of ionized regions 
$Q_{\rm HII}(z)$
in the left panel of Figure \ref{Q_z}. It is clear that the
two models differ only at $z > 6$; for the LRM
$Q_{\rm HII}$ evolves from 0.7 to unity in the redshift range 
6.6--6.0, implying that the universe is in the pre-overlap stage, while 
for the ERM, $Q_{\rm HII} = 1$ at these epochs.

We next compare the 
evolution of the globally volume-averaged photoionization rate
for the two models
in the middle panel of Figure \ref{Q_z}. At $z<6$ the ionizing sources are 
mainly PopII stars and QSOs for both models and so $\Gamma_{\rm HI}$ is 
comparable 
in the two models. At $z>6$, there 
are no contributions from PopIII stars to the UV background radiation
in the LRM, 
which are instead present in the ERM. As a consequence, at high redshift, 
$\Gamma_{\rm HI}$ is higher in the ERM with 
respect to the late reionization one.
The photoionization rate evolution is also 
compared with the results obtained by F02 and \citeN{bhvs05}, hereafter 
B05. The agreement is quite good with 
B05 data at $z \le 4$; the mild 
discrepancy at $z \approx 4$ could be attributed to the 
systematic overestimation of the photoionization rate due to 
the limited box size and 
resolution of their hydrodynamical simulations.
On the other hand, the ERM mildly violates the 
upper limit of $\Gamma_{\rm HI} = 0.08$ at $z=6.0$ obtained by F02. 
However F02 use $\Gamma_{\rm HI}$ as a free parameter (same as in  B05) 
to match the mean transmitted flux at high redshift, whose 
value has a large uncertainty (see Section \ref{mtf}). So we 
expect that also the values of $\Gamma_{\rm HI}$ at $z>4$ are associated with 
uncertainties at least as large as for the estimates at lower redshifts. 
This alleviates the mild discrepancy between the ERM and the upper limit on 
$\Gamma_{\rm HI}$ suggested by F02 at $z=6$.

The globally volume averaged neutral hydrogen fraction $f_{\rm HI}$ is shown 
in the right panel of Figure 
\ref{Q_z}. The evolution of $f_{\rm HI}$ with redshift has been obtained 
computing 10 LOS for eight redshift bins ($\Delta z=0.4$) covering the 
redshift interval 2.8-6.7. 
Figure \ref{Q_z} shows that, as expected, the IGM is highly ionized
for the ERM, while it is quite neutral 
at $z \gtrsim 6$ for the LRM. Furthermore, and opposite to the ERM case, 
we find a sharp 
evolution in the 
neutral fraction around $5.5 < z < 6.5$ for the LRM when overlapping occurs 
in that model. Our prediction on the neutral hydrogen fraction at $z=6$, 
considering cosmic variance, is in agreement with the result obtained by F02, 
while there is a discrepancy with the measure of $f_{\rm HI}$ arising from the 
analyses of the HII regions \cite{wlc05,wl04,mh04} 
as discussed in the Introduction.

\subsection{Additional physics}

\begin{figure}
\centerline{\psfig{figure=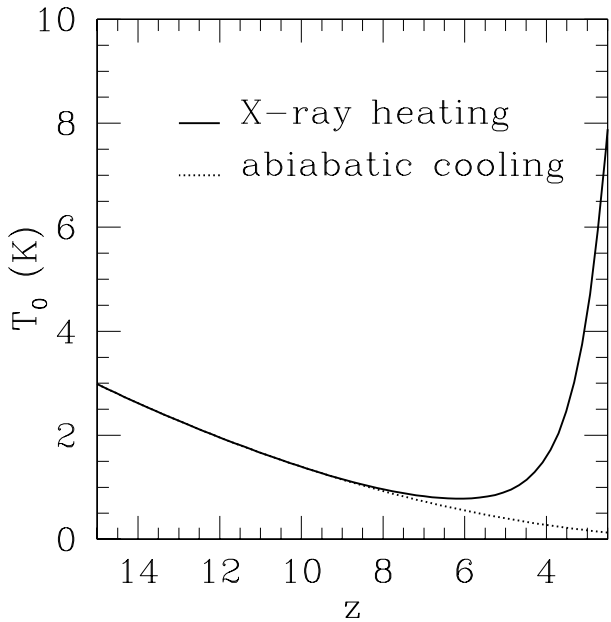,width=8.0cm,angle=0}}
\caption{Temperature evolution of the neutral regions. The solid
line refers to the model with X-ray heating described in the text; the 
dotted line describes the temperature evolution  
assuming adiabatic cooling.}  
\label{tneutral}
\end{figure}

In the two-phase model, the photoionization rate $\Gamma_{\rm HI}$
is nearly zero in the neutral HI regions, as most of the points 
are opaque to ionizing radiation. 
On the other hand, the photoionization rate
$\Gamma_{\rm HI}^{\rm HII}(z)$ inside HII regions  
is assumed to be homogeneous; a natural way to relate it to
the globally volume-averaged photoionization
rate $\Gamma_{\rm HI}(z)$ is through the relation:
$\Gamma_{\rm HI}^{\rm HII}(z) = \Gamma_{\rm HI}(z)/Q_{\rm HII}(z)$.

Also to be noted is that 
in absence of ionizing radiation, the temperature $T_0$ of the 
mean density neutral gas will decrease adiabatically and can be 
as low as $\sim 1$ K at $z \approx 6$;
however, the presence of a population of hard
photons (say, soft X-ray photons from QSOs, X-ray binaries or
supernova remnants), which are able to penetrate the atomic medium, 
can raise $T_0$ for these neutral regions \cite{cm04}. To investigate this 
possibility we study the evolution of $T_0$ for neutral regions in the 
expanding Universe, integrating the following equation:
\begin{equation}
(1+z)\frac{dT_0}{dz}=2T_0-\frac{2}{3}
\frac{\Gamma_{\rm tot}-\Lambda_{\rm tot}}{H(z) n_b k_B}
\end{equation}
where $\Gamma_{\rm tot}$ and $\Lambda_{\rm tot}$ 
are the total heating and cooling 
rates, respectively.
Assuming that the main heating mechanism is due to soft X-ray photons, 
$\Gamma_{tot}$ can be substituted by $\Gamma_X$, where $\Gamma_X$ is the 
X-ray heating rate. Following \citeN{cm04}, we parametrize 
the emissivity in X-rays in terms of the fraction of energy that is emitted in
X-rays compared to the energy emitted at Ly$\alpha$ per unit log $\nu$, 
which we 
designate as $\alpha_X$. The X-ray heating rate is then given by the following 
relation:
\begin{equation}\label{tevol}
\Gamma_X(z)=0.14 ~ \alpha_X ~ h_{\rm P} ~ \nu_{\alpha}~ \epsilon(z)
\end{equation} 
where $0.14$ is the fraction of the X-ray energy used to heat the gas 
\cite{sv85}, $h_{\rm P}$ is 
the Planck constant, $\nu_{\alpha}$ is the frequency of the Ly$\alpha$ 
transition
and $\epsilon$ is the comoving Ly$\alpha$ emissivity as obtained in the LRM.

We start solving the differential equation (\ref{tevol}) 
from $z_{\rm start}=30$ assuming 
that, at this redshift $\epsilon(z) \approx 0$. 
In the absence of any heating sources, 
the temperature of the gas can be shown to be $\sim$ 11K. 
\footnote{The Compton scattering between the 
CMB photons and relic free electrons from cosmic 
recombination couples the cosmic gas temperature to the CMB one, down to 
redshift $1+z_f\sim 1000(\Omega_b h^2)^{2/5}$ (\cite{peebles93}). 
Following that the 
temperature of the gas cools down adiabatically. So the gas temperature
before the formation of any heating source is given by
$T_0(z_{\rm start})=T_{\rm CMB} (1+z_f)
\left[(1+z_{\rm start})/(1+z_f)\right]^2$, 
where $T_{\rm CMB}$ is the temperature of the CMB at $z=0$.}
During the redshift range covered in the calculation, 
the temperature of the gas
is low enough to neglect cooling as the cooling function is different 
from zero only for temperatures above $10^4$ K.

Figure (\ref{tneutral}) shows the thermal evolution for neutral regions 
when we include X-ray heating with $\alpha_X=0.01$. Assuming that only 
1 per cent of 
the energy is emitted as X-rays, the temperature of the gas is equal to 0.76 K 
at $z=6$. Even when we use a higher value 
of $\alpha_X=0.1$, the temperature raises only to 2.8 K at $z = 6$.

Fortunately for this work, it
turns out that {\it all} the results are independent of the 
precise value of $T_0$ for neutral regions as long as it is below 
1500 K,
with variations being less
than the typical statistical variance. 
The reason for the insensitivity of our results to $T_0$ can be understood 
in the following way: the value of $T_0$ has two possible effects
on the simulated spectra. The first effect is to determine the recombination
rate of the ionized species and thus affect the neutral hydrogen fraction. 
However, the low ionization rates in the neutral regions 
imply very small ionized fraction and thus a very long recombination time, 
thus making the value of $T_0$ irrelevant for 
calculating the neutral fraction. The second effect of $T_0$
is to determine the widths of the lines through the Doppler profile. 
But here again, because of the large neutral fractions, the line profile is
predominantly determined by the natural width (which shall be discussed 
in detail later) and thus the 
effect of $T_0$ is again negligible.
For the rest of the paper, we 
shall assume that the temperature $T_0$ for neutral regions is 1K.

There are few more subtleties which need to be addressed while 
dealing with neutral regions along lines of sight. First,
the volume filling 
factor $Q_{\rm HII}(z)$ applies to three dimensional regions 
only, and hence one needs to translate this into 
a one dimensional filling factor $q_{\rm HII}(z)$
along different lines of sight 
in a consistent manner which takes into account the evolution 
in $Q_{\rm HII}$.
This is a purely geometrical exercise and can
be performed if one knows the geometry of the neutral regions.
However, the value of the filling factor $Q_{\rm HII}$ does not uniquely 
determine the size and shape of the neutral regions; the detailed 
topology depends
on the nature of sources, coupled with the 
density distribution of the IGM, and hence is non-trivial
to take it into account analytically. On the other hand, 
numerical simulations still do not have enough dynamic range
to address this issue for wide regions of parameter space. 
Given this, we devise an approximate method, described below, to calculate 
$q_{\rm HII}(z)$ along different lines of sight. In addition, we also
study different variations of our method accounting for different 
topologies of the neutral regions, and check whether 
our main conclusions remain unchanged.

The simplest method of distributing the neutral regions along 
different lines of sight is based on the assumption that 
the positions of the neutral regions in the three-dimensions are completely 
random and the regions are {\it not} correlated with the density field.
This assumption seems to be quite reasonable at late stages of reionization
as found in radiative transfer simulations, 
where most of the individual ionized regions have overlapped leaving
neutral regions of random shapes and sizes with no significant clustering
pattern. 
(For a visual impression see, for example, the maps in \citeNP{cfw03}.)
Of course, very high density regions, like
collapsed structures or filaments, tend to remain neutral till late times, thus
correlating the neutral regions with density field. However, 
these high density regions are not significant for the Ly$\alpha$ forest, and 
hence can be ignored in our analysis. Nevertheless, we do check 
the effects of clustering of 
neutral regions of large sizes and their correlation 
with the density field in Section \ref{variations}. 
In addition we present the technical details
of distributing the neutral regions along lines of sight and their
physical properties in Appendix 
\ref{q_hii}.

\section{Results}
\label{res}
\begin{figure*}
\psfig{figure=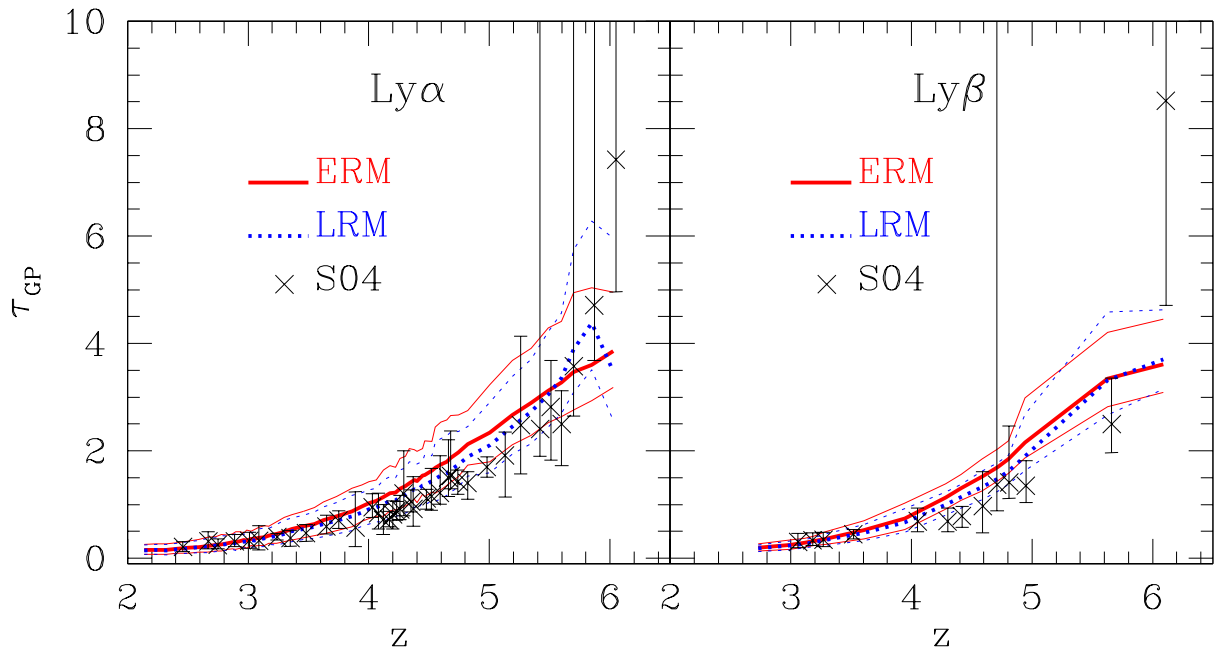,width=16.0cm,angle=0}
\caption{Ly$\alpha$ (left panel) and Ly$\beta$ (right panel) GP optical depth 
compared with data from Songaila 
(2004). Solid (red) and dotted (blue) thick lines represent average results 
for ERM and LRM, 
respectively, on 100 LOS for each 
emission redshift; the thin lines denote the upper and lower transmission 
extremes in each bin, weighted on 100 LOS.} 
\label{flux}
\end{figure*}

In this Section, we present the main results of our analysis. In 
the first part, we shall put our model to test by comparing it
with various available observations at $z < 6$. As we shall see,
the model is quite successful in matching the observational data. 
Next, we shall discuss the predictions of the model 
at $z > 6$ in the second part, and then try to determine the ionization state
of the IGM.

\subsection{Comparison of the model with observations at $z < 6$}

\subsubsection{Gunn-Peterson optical depth ($\tau_{\rm GP}$)}
\label{mtf}

The first obvious test for our model would be to check whether it can match 
the mean GP opacity of the Ly$\alpha$ and Ly$\beta$ forests at $z < 6$. 
For this purpose, we use the data from S04, and thus
the procedure for obtaining the mean transmitted flux 
from simulated spectra is similar to that work.
We consider the emission redshifts of the 50 QSOs observed 
($2.31<z_{\rm em}<6.39$)
as mentioned in Tables 1 and 2 in S04. 
For each emission redshift, we simulate the Ly$\alpha$ absorption spectra 
covering the wavelength range 1080 - 1185 \AA. We then divide each spectra 
in seven parts of length 15 \AA \ and compute the Ly$\alpha$ 
Mean Transmitted Flux (MTF)
for each part. These data points are then 
binned in a way such that each bin contains six points.
For each bin, the mean transmission and the 
extremes of transmission have been computed and then assigned to the median 
redshift within the bin. The GP optical depth is defined as: 
$\tau_{\rm GP}=-\ln({\rm MTF})$.
The GP optical depth evolution for Ly$\alpha$ transition 
is plotted in Figure \ref{flux} (left panel) as a function of the median 
redshift in 
each bin. The vertical 
error bars show the range of extremes of transmission within each bin, 
weighted on 100 LOS, translated to optical depth.

In order to compare our results with observational ones 
(S04), each spectra have been convolved with a Gaussian whose FWHM is equal to 
8~km~s$^{-1}$, if the emission redshift $z_{\rm em}$ is below 4, 
or equal to 56~km~s$^{-1}$, if $z_{\rm em}>4$. 
This procedure implies, in the first case, a resolution around 36000, 
since the observed spectra have been taken with the HIRES spectrograph 
on the Keck I telescope, while in the second case the mimicked resolution 
is 5300, the same of the ESI spectrograph on the Keck II telescope. 
The pixel size of the rebinning is 12~km~s$^{-1}$. 
  
Extending the above procedure for Ly$\beta$ region of the absorption
spectra (corresponding to the rest wavelength range 980 - 1010 \AA), 
we derive
the \emph{total} 
optical depth at a given redshift $z$ from the sum of the direct 
Ly$\beta$ absorption at that redshift 
and the Ly$\alpha$ absorption at redshift 
$1+z_{\beta}=\frac{\lambda_{\beta}}{\lambda_{\alpha}}(1+z)$, i.e.,
\begin{equation}
\tau^{\rm tot}_{{\rm Ly}\beta}(z)=\tau_{{\rm Ly}\beta}(z)
+\tau_{{\rm Ly}\alpha}(z_{\beta}),
\label{lybeta}
\end{equation}
where $\tau_{{\rm Ly}\alpha}$ and $\tau_{{\rm Ly}\beta}$ are given 
by equations (\ref{lyalphatau}) and (\ref{lybetatau}) respectively.
For each emission redshift, 
the absorption spectra in the wavelength range 980 - 1010 \AA \  
is divided in two parts of 30 \AA \ each.
As in the Ly$\alpha$, we obtain the evolution of Ly$\beta$ MTF
and the range of extreme values by binning the data.
Note that, in order to calculate the Ly$\beta$ flux distribution in the rest 
wavelength range 980 - 1010 \AA \  (as discussed above), we need to 
estimate the Ly$\alpha$ optical depth in the interval 827 - 1010 \AA \ [this 
follows trivially from the expression (\ref{lybeta}) for 
Ly$\beta$ optical depth]. 
The evolution of the MTF for Ly$\beta$ is plotted in  
Figure \ref{flux} (right panel). The error bars show the range of extremes of 
transmission within each bin, translated to optical depth.

We find that both the Ly$\alpha$ and the Ly$\beta$ MTFs are in excellent 
agreement\footnote{The fact that there is hardly any 
difference between the early and late reionization models is related to the 
fact that the two models differ substantially only at redshifts above 6.} with 
observations at $3 \lesssim z \lesssim 6$. 
Though it might seem 
from Figure \ref{flux} that it is 
difficult to reach an optical depth as high as S04 at 
$z=6$ with our models, it must be noted that our results 
are weighed over a large number 
of realizations. In fact, we find that there are lot of realizations 
which give very high optical depth at $z=6$, in complete agreement with S04.

Note that in the MTF analysis of the Ly$\alpha$ forest, the slope of the 
equation of state $\gamma -1$ has been treated as a (non-evolving) 
free parameter which has the best-fit 
value $\gamma=1.3$. Once that it is fixed in order to match the 
Ly$\alpha$ data, no any other free parameter can be tuned to 
obtain a good trend of Ly$\beta$ MTF. 
The agreement of the MTFs in both the Ly$\alpha$ and Ly$\beta$ 
regions is thus an indirect confirmation that the lognormal model for density 
fields could be considered as a fair description of 
the mildly non linear regime.

\subsubsection{Probability Distribution Function (PDF)}
\label{pdf}

\begin{figure*}
\psfig{figure=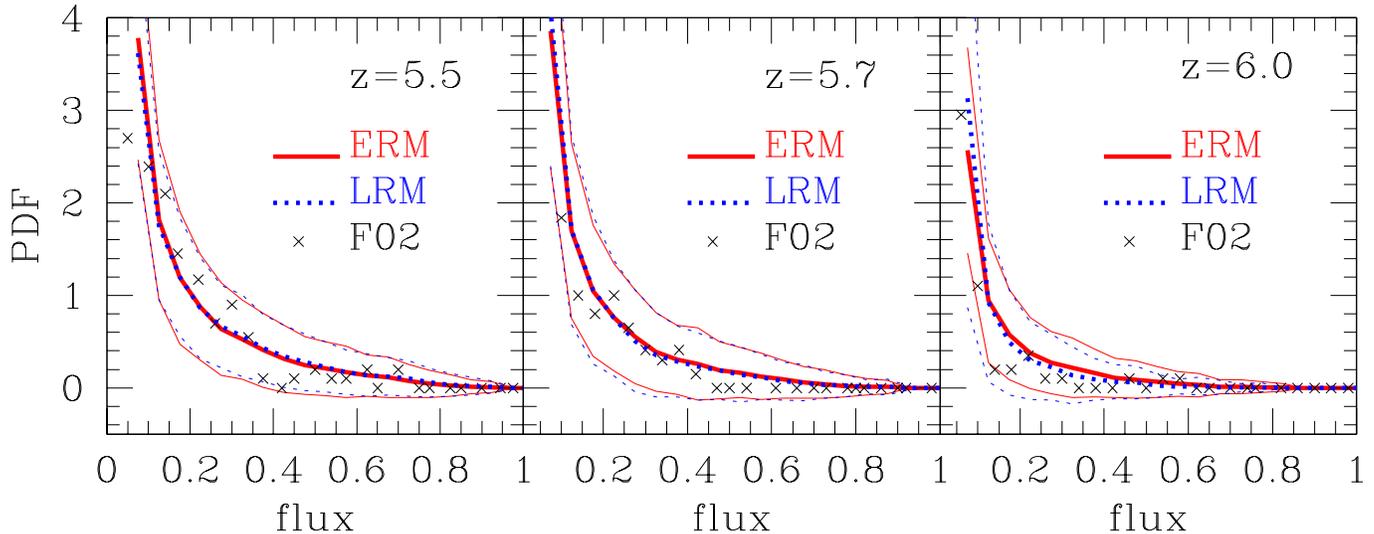,width=18.0cm,angle=0}
\caption{Probability distribution function of the 
transmitted flux at mean redshift 5.5, 5.7 and 6.0, respectively, compared 
with F02 Keck data. Solid (red)
and dotted (blue) lines represent the ERM and the LRM, respectively. 
For each model, the thick 
line is the average over 500 LOS, the thin lines denote the 
cosmic variance.} 
\label{pdflowz}
\end{figure*} 

In this Section, we compute the Probability Distribution Function
of the transmitted flux for the Ly$\alpha$ forest and compare it
with the observed Keck spectra of 
SDSS 1044-0125, SDSS 1306+0356, and SDSS 1030+0524 within redshift
range $5.5 < z < 6.0$ (F02).
In order to compare with observations, we add the relevant observational 
artifacts in our simulated
spectra, i.e., we smooth the simulated spectra with a Gaussian filter of 
smoothing length $\sigma_v$ = 28~km~s$^{-1}$ 
(corresponding at a FWHM of 66~km~s$^{-1}$) and bin them in pixels of 
width 35~km~s$^{-1}$. We then add noise (see Section 2.1.5) to the simulated 
Ly$\alpha$ forest 
spectra  corresponding to the observed data with $\sigma_{\rm noise} = 0.02$.
Figure \ref{pdflowz} shows the PDF of the 
transmitted flux, computed using 500 random realizations of the 
artificial spectra at the mean redshift 5.5, 5.7, 6.0 respectively. 

The flux PDF of the simulated spectra is consistent with 
the observational ones in all three redshift cases. Furthermore, as expected, 
the 
agreement is quite good for both reionization models. 
Note that the difference in the photoionization rates of the two models are 
typically $18\%$, $18\%$ and $39\%$ for redshifts 5.5, 5.7, 6.0 
respectively (see middle panel of Figure \ref{Q_z}); 
yet the two models seem to be indistinguishable.
This implies that the PDF is not sensitive enough in discriminating between 
different evolution histories of the ionizing background.

It is worth briefly mentioning here that 
F02 have used numerical simulations 
to compute the absorption spectra of high-redshift quasars in the 
Ly$\alpha$ region and  have found a rapid evolution of the 
volume-averaged neutral fraction of hydrogen  at $z \leq 6$
(from $f_{\rm HI} \sim 10^{-5}$ at $z = 3$ to 
$f_{\rm HI} \sim 10^{-3}$ at $z = 6$). 
This evolution has been interpreted as a signature of the end of reionization
around $z \approx 6$. However, we find that, in addition to late 
reionization models 
(where $f_{\rm HI}$ is  evolving rapidly at $z \approx 6$), early 
reionization models also give a good fit to the observed MTF and PDF at 
$z\leq 6$. It is thus {\it not} possible to rule out 
early reionization models using only MTF and PDF statistics at $z \lesssim 6$.

\subsubsection{Dark Gap Width Distribution (DGWD)}

\begin{figure*}
\centerline{\psfig{figure=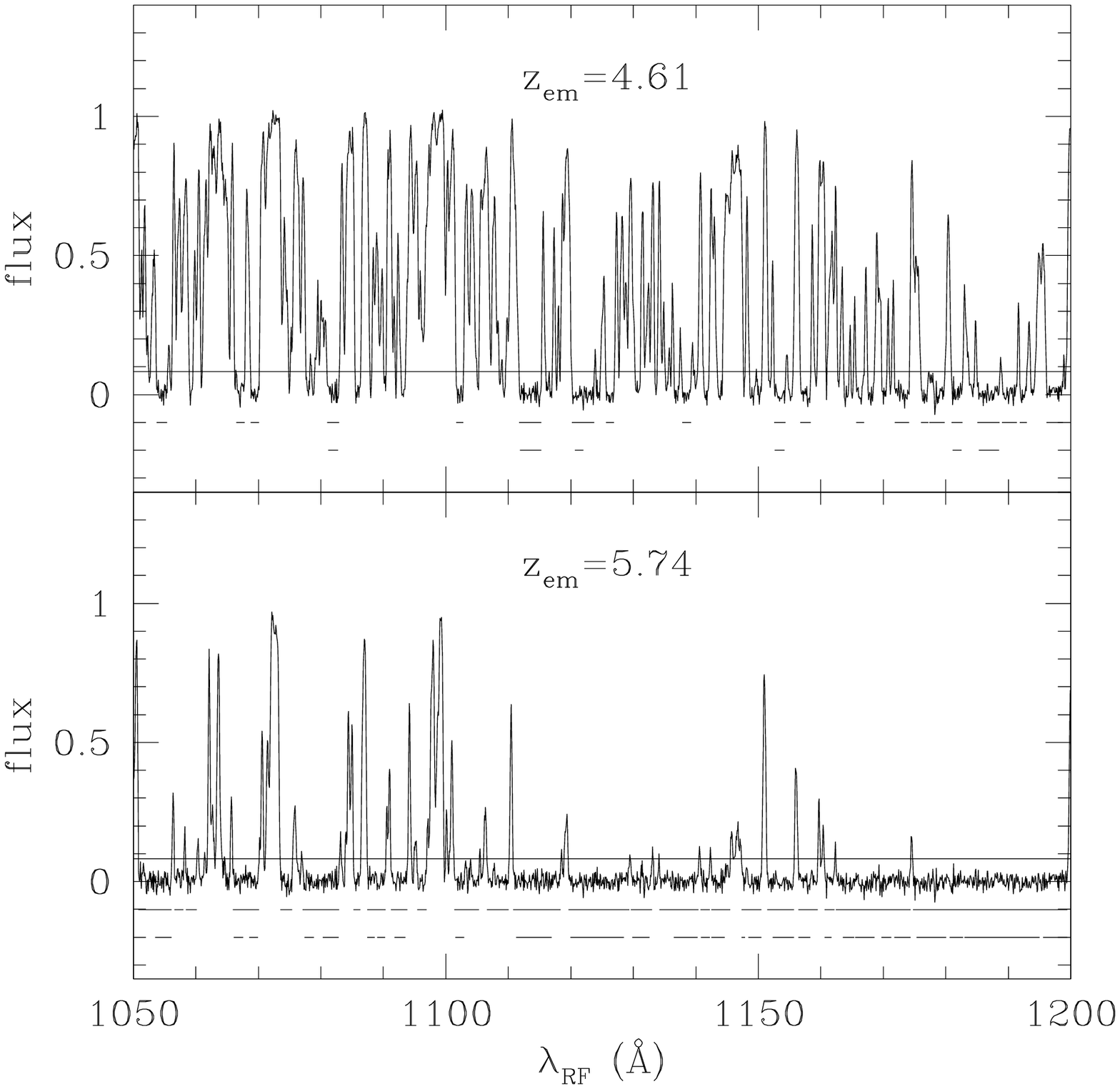,width=16.0cm,angle=0}}
\caption{Two simulated QSO spectra with emission redshift 4.61 (top panel) 
and 5.74 (bottom panel). 
The broken lines immediately below the spectra represents dark gaps in 
the Ly$\alpha$ region and the ones 
below those show the analogous gaps in the Ly$\beta$ region.
The horizontal solid lines 
slightly above zero transmission represent the flux threshold ($=0.082$)
used for defining gaps.} 
\label{2spc_gaps}
\end{figure*} 

\begin{figure*}
\centerline{\psfig{figure=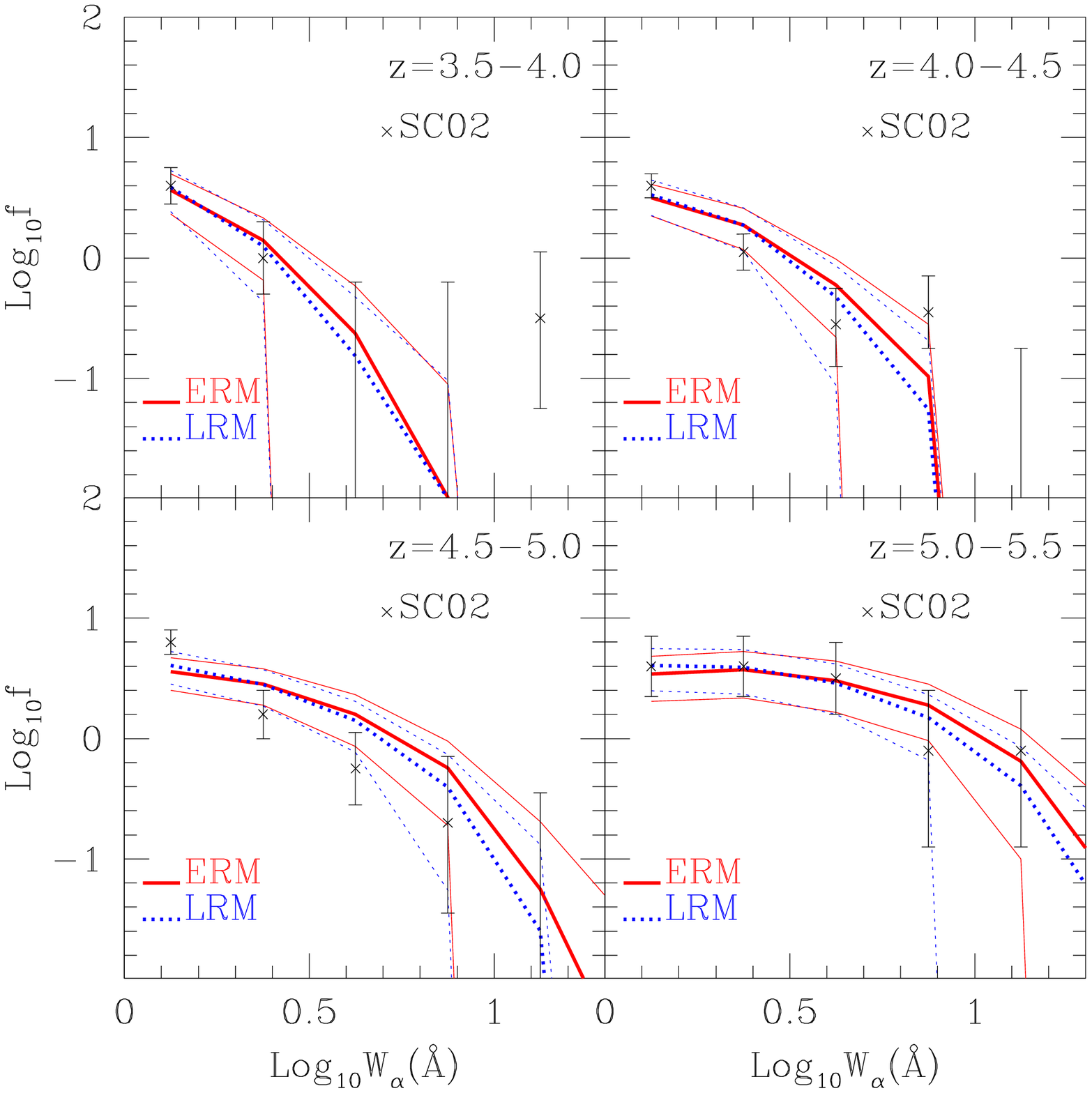,width=16.0cm,angle=0}}
\caption{Dark Gap Width Distribution (DGWD) at redshift 3.5--5.5 compared with 
Songaila \& Cowie (2002), 
denoted SC02 in the panels. Solid (red)
and dotted (blue) lines represent the ERM and the LRM, respectively. 
For each model, the thick 
line is the average over 300 LOS, the thin lines denote 
cosmic variance.} 
\label{dgwd55}
\end{figure*} 

At high redshifts, regions with high transmission in the Ly$\alpha$ forest 
become rare. Therefore an alternative method to analyze the 
statistical properties of the transmitted flux is the distribution 
of dark gaps first suggested by \citeN{croft98}, defined as contiguous regions
of the spectrum having an optical depth $>$ than 2.5 over rest frame 
wavelength intervals greater than 1 \AA.
In this Section we will compare our results with observational data 
obtained by \citeN{sc02}, hereafter SC02, analyzing 15 high-redshift 
QSOs whose emission redshifts lie in $4.42\lesssim z\lesssim 5.75$. 
In order to obtain a fair comparison with data, each simulated spectrum 
has been convolved with a Gaussian having FWHM equal to 60~km~s$^{-1}$, 
resulting in a spatial resolution similar to the data obtained using 
ESI spectrograph.   
In Figure \ref{2spc_gaps} we plot  typical simulated line of 
sight spectra at redshifts 4.61 and 5.74. The black lines 
plotted immediately below the spectra show the regions 
identified as gaps in the Ly$\alpha$ region and the ones 
below those show the gaps present in the Ly$\beta$ region too.
This Figure should be compared with Fig 2
of SC02, showing good qualitative agreement between our results and 
observations. It is also clear from the Figure that 
the frequency and the width of the gaps increase 
from redshift 4.61 to redshift 5.74.

Figure \ref{dgwd55} shows the DGWD
for the redshift range $3.0<z_{\rm abs}<5.5$ (where $z_{\rm abs}$ is 
the redshift of the absorber), obtained from a sample of 300 LOS for each 
redshift bin.
 The distribution essentially measures the number of gaps having a 
certain width $W_{\alpha}$ in the QSO rest frame
within the Ly$\alpha$ region.
The results, as well as the statistical errors, obtained from our models 
are in good agreement with observational data over a wide redshift range. 
One can also see that the frequency of larger gaps increases as we go
to higher redshifts. Comparing our results with hydrodynamical simulations
\cite{pn05} we find that our results are in 
better agreement with observations. This is probably due to their limited 
box size (6.8 comoving $h^{-1}$Mpc); in this case to simulate spectra covering 
a
 large redshift range (5.0--5.5, for instance) each line of sight has to cross
the simulated volume more than once. So the presence of an under-ionized region
could break a long dark gap in smaller ones. The advantages of using 
semi-analytical simulation is that we can obtain the same length of the 
observed 
spectrum (e.g. 100 \AA \ to compare with SC02) in one 
realization, instead of combining together various artificial spectra end to 
end. 

So far we have tested our models against the observational data 
available at $z < 6$. We have seen that both late and early reionization 
models are able to match (i) the MTF evolution both in the Ly$\alpha$ or 
Ly$\beta$ regions, (ii) the PDF of the transmitted flux and 
finally (iii) the DGWD. We can thus conclude that the results obtained 
at $z < 6$ 
do not allow to exclude the possibility that the universe has been 
reionized as early as at redshift 14. 
However, we have already discussed the fact that the difference between
the two reionization scenarios are most substantial only 
at $z > 6$. It is thus important to see how 
the Ly$\alpha$ forest at $z > 6$ can be used for distinguishing between 
the two different scenarios.
This is what will be done in the next Section.

\subsection{Predictions for higher redshifts}

\begin{figure*}
\psfig{figure=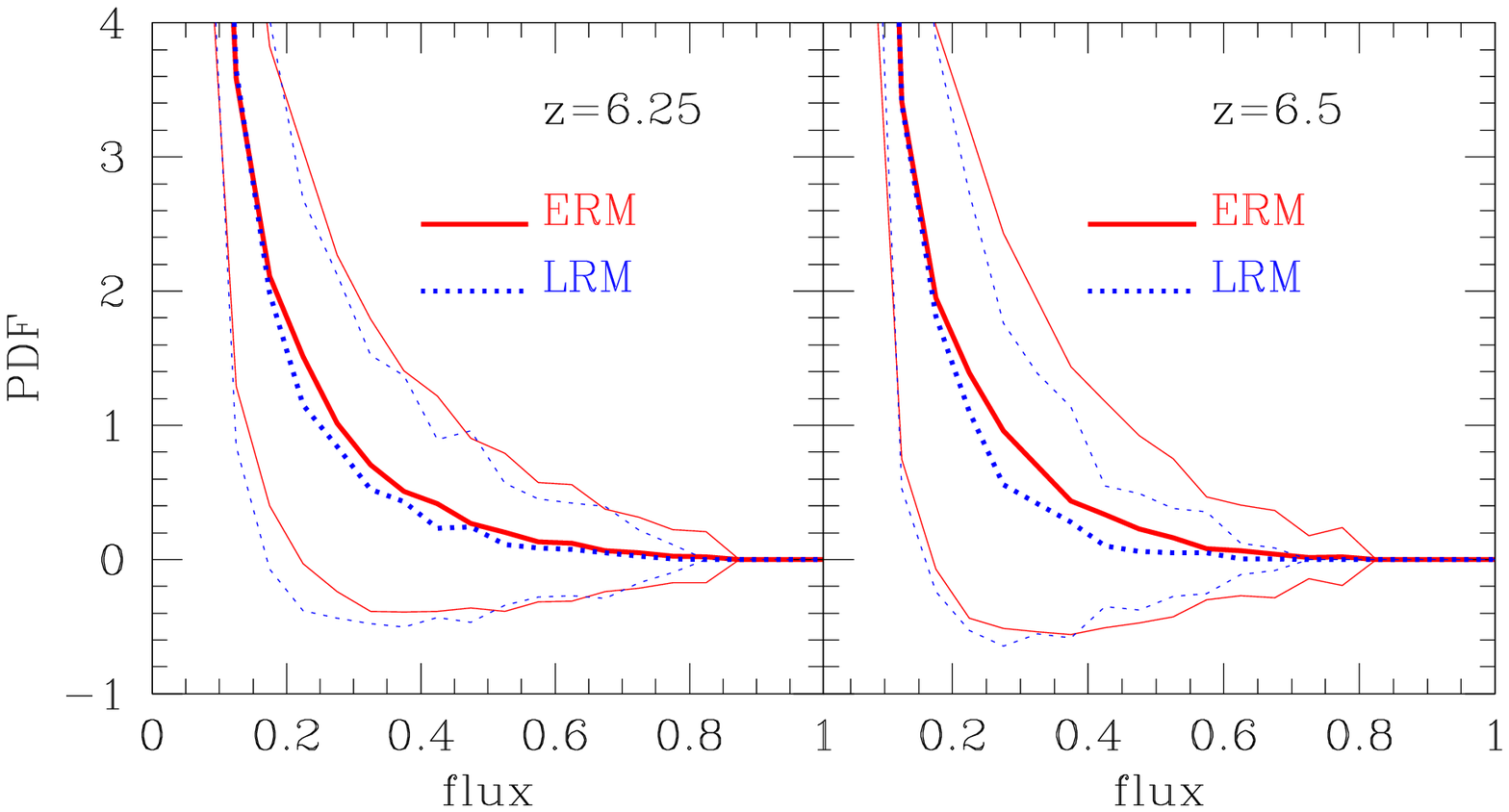,width=16.0cm,angle=0}
\caption{Same as in Figure \ref{pdflowz} but at mean redshifts 6.25 and 6.5.} 
\label{pdf6}
\end{figure*} 

\begin{figure*}
\centerline{\psfig{figure=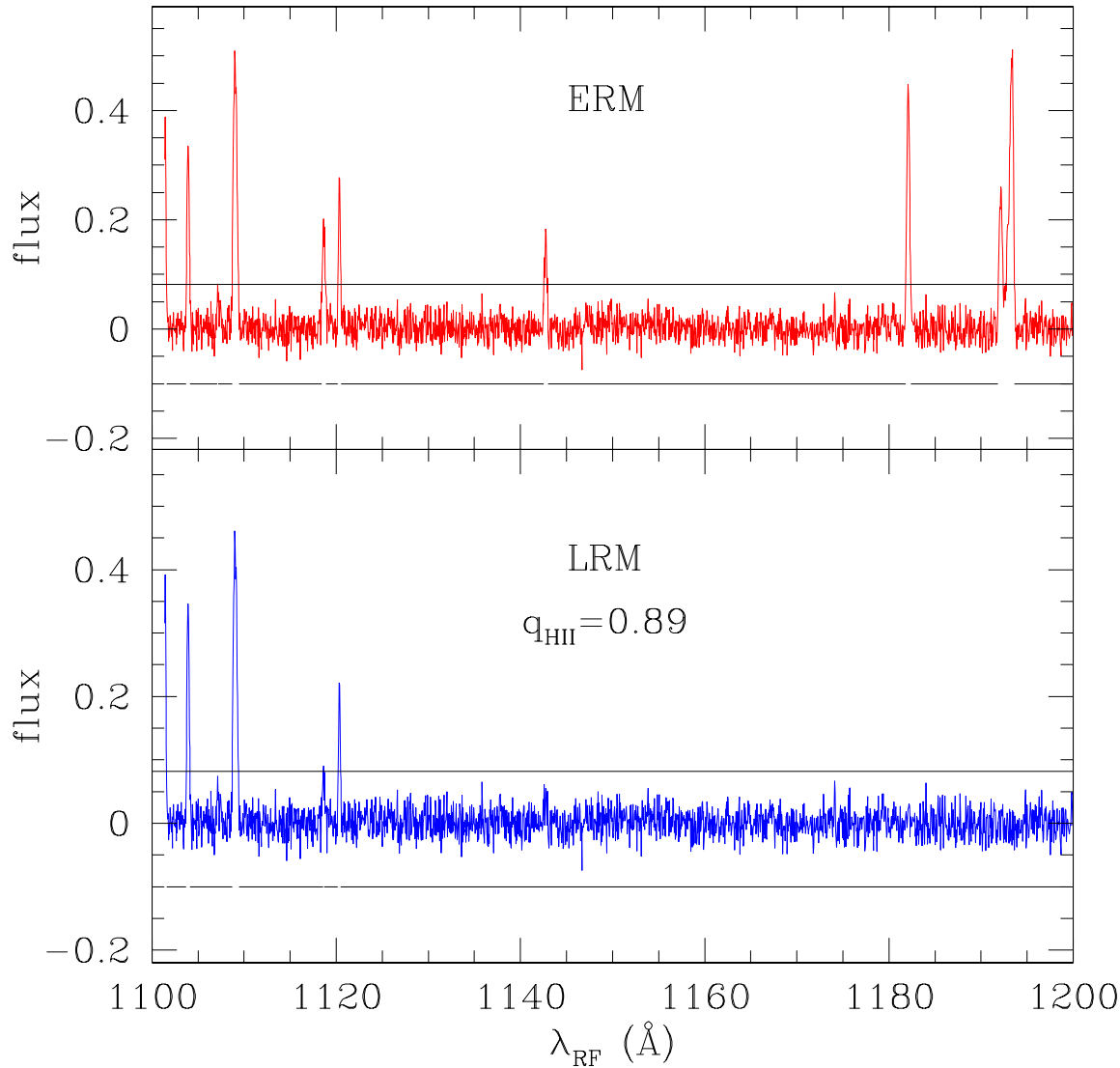,width=16.0cm,angle=0}}
\caption{Comparison between two simulated spectra for different
reionization scenarios, ERM (top panel) and LRM (bottom panel), 
in the redshift range 
5.7--6.3. Both these spectra are obtained from the same 
density distribution.} 
\label{largest1}
\end{figure*} 

\begin{figure*}
\centerline{\psfig{figure=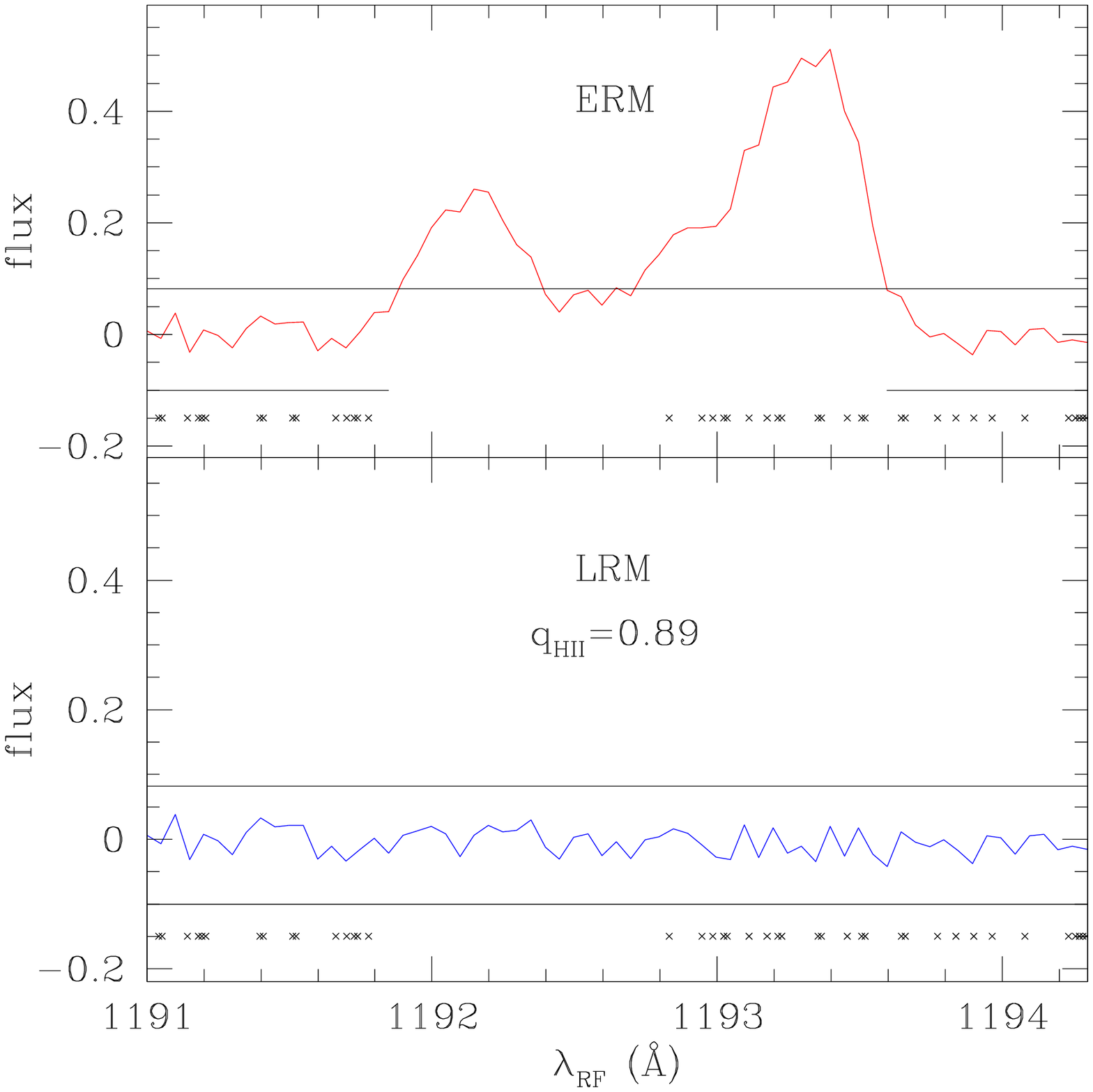,width=16.0cm,angle=0}}
\caption{Same as in Figure \ref{largest1}, but zoomed into the spectral region 
$1191$ \AA\  $< \lambda_{\rm RF} < 1194$ \AA. The black lines 
plotted immediately below the spectra show the regions 
identified as gaps.
We have also
shown the positions of the neutral pixels by crosses. Even if neutral pixels 
are present only in the LRM, we draw them also in the ERM spectrum, in order 
to visualize their location. 
It is evident that 
there is suppression of 
the flux in the LRM even if there are no corresponding 
neutral pixels. This result, as 
discussed in detail in the text, is due to the damping wings of the 
neighboring neutral pixels.} 
\label{zoom}
\end{figure*}

Let us now extend the analyses of the previous Section to 
spectra at $z > 6$ and try to determine whether the Ly$\alpha$
forest is able to distinguish between different models of 
reionization. Since the spectra are generally expected to be much 
darker at these high redshifts, it is clear that the MTF would not 
be able to distinguish between the two models (it is 
consistent with zero irrespective of the ionization
state of the IGM). Hence, we start our discussion
with the PDF of the transmitted flux for the Ly$\alpha$ forest.

\subsubsection{Probability distribution function (PDF)}

As in the Section \ref{pdf}, we compute the PDF of the transmitted 
flux for 500 random LOS, 
with the corresponding cosmic variance. The results are shown in 
Figure \ref{pdf6} for mean redshifts of 6.25 and 6.5, respectively.   
Interestingly, we find that even 
at redshifts higher than 6 (where the ERM and LRM differ quite substantially 
in their physical properties), the PDF is not able to differentiate 
between the two reionization scenarios mainly because of large 
cosmic variance. This has to do with the fact that most of the pixels 
have flux consistent with zero for both reionization models and so 
the PDF essentially probes the noise 
distribution (which is independent of the physical state of the IGM). However,
note that in the LRM we have {\it no} pixels with $F > 0.8$ at $z = 6.25$
and with $F > 0.7$ at $z = 6.5$ respectively, while there are pixels
(though very few)
with $F$ as high as 0.85 at $z = 6.25$ and 0.8 at $z = 6.5$ for the ERM 
(these correspond to some peaks present in the ERM which are suppressed 
in the LRM). Whether this can discriminate between the two models is doubtful 
particularly because of the uncertainties in the continuum of the 
unabsorbed quasar spectra and the effects arising from atmospheric absorption.

\subsubsection{Dark Gap Width Distribution (DGWD) at $z > 5.5$}

The next statistics which can be used is the DGWD for the 
Ly$\alpha$ forest  at redshifts higher then 5.5. 
For definiteness, we consider two redshift intervals: 
5.7 -- 6.3 and 6.0 -- 6.6. The first case should be 
applicable to QSOs having emission redshift around 6.4, while 
the second case corresponds to an emission redshift $\approx$ 
6.7.
In Figure \ref{largest1}, we plot 
sample spectra for the two different reionization models 
in the redshift range 5.7 -- 6.3, with the upper (lower) 
panel corresponding to ERM (LRM). The black lines 
plotted immediately below the spectra show the regions 
identified as gaps.
Note that the two spectra have the same baryonic density distribution
and differ only in the distribution of neutral regions.
It is clear that at 
large values of rest frame wavelengths ($\lambda_{RF}$), say, 
$\lambda_{\rm RF} > 1150$ \AA \ 
(which corresponds to $z > 6$), there are
substantial differences between the ERM and LRM. The 
LRM does not have the peaks at large redshifts which are present in the 
ERM, and hence one obtains gaps of much larger 
widths for the LRM. The reason for the suppression 
of the peaks in the LRM is twofold.
Firstly there is an increase in the optical depth at the pixels 
where neutral regions are placed, thus decreasing the transmission.
However, there is a second effect which seems to be more important which
has to do with the damping profile of neutral hydrogen arising 
from natural line width. This effect can suppress flux in regions
which are not necessarily neutral but lie in the vicinity of a 
highly neutral region. 
This can be understood
from a close-up of the spectra shown in Figure \ref{zoom} where 
we have zoomed into a region between 
$1191$ \AA\  $< \lambda_{\rm RF} < 1194$ \AA. 
As before the upper (lower) 
panel corresponds to ERM (LRM) and the black lines 
plotted immediately below the spectra show the regions 
identified as gaps.
We have also
shown the positions of the neutral pixels by crosses within the Figure.
Note that there are two prominent peaks in the ERM at
$\lambda_{\rm RF} = 1192.2$ \AA \ and $1193.4$ \AA \ respectively while they
are completely suppressed in the LRM. However, note that there are
{\it no} neutral pixels at the location of the peaks -- they
are actually suppressed by the damping wings of a small number of 
neutral pixels in the vicinity. Thus the damping wing of 
neutral regions can have a dramatic effect on the distribution of 
dark gap widths as shall be discussed next.

\begin{figure*}
\psfig{figure=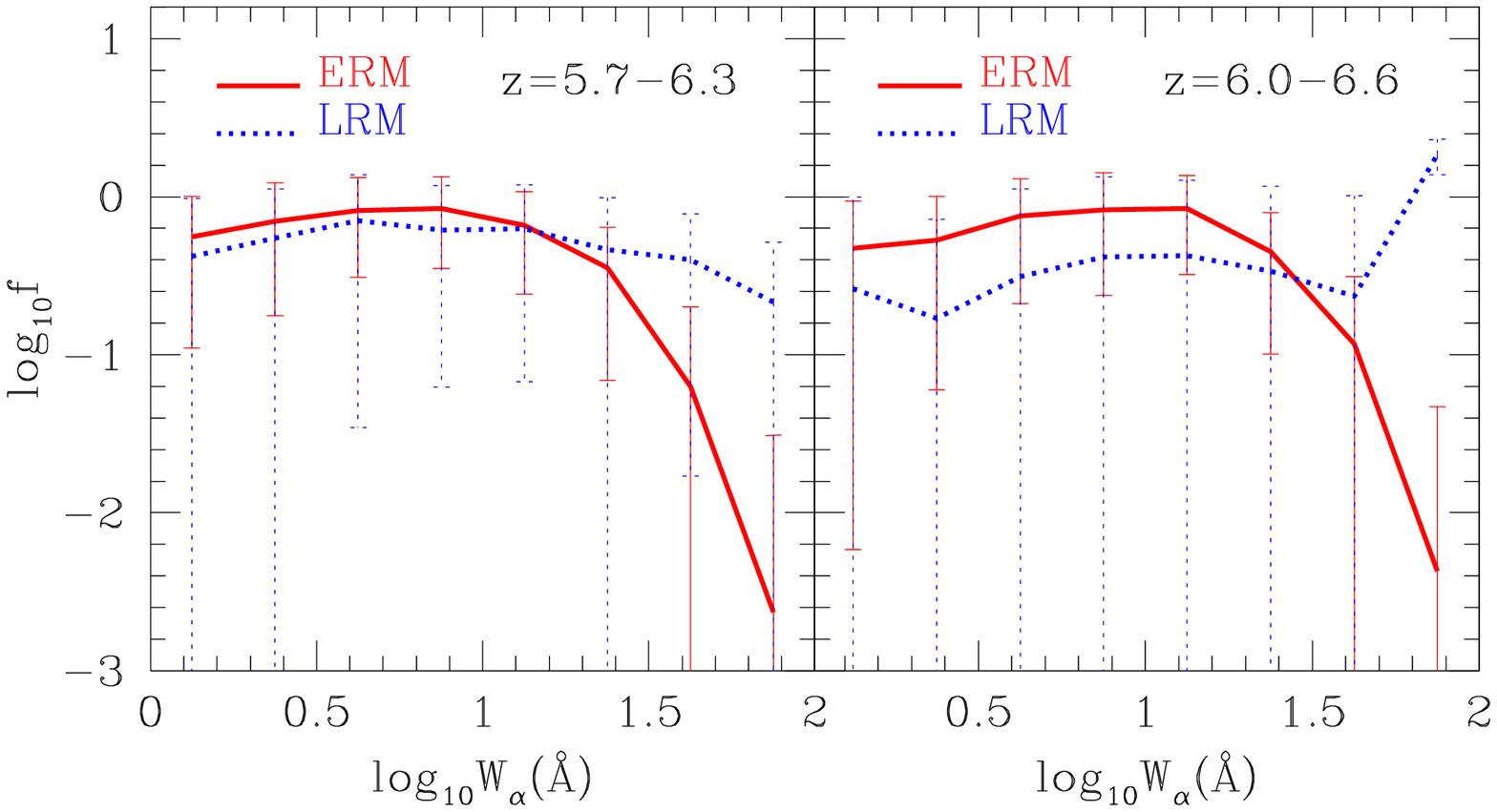,width=16.0cm,angle=0}
\caption{Dark Gap Width Distribution (DGWD) at redshift  5.7--6.3 (left) and 
6.0--6.6 (right). Solid (red)
and dotted (blue) lines represent the ERM and the LRM, respectively. 
For each model, the thick 
line is the average over 300 LOS, the error bars denote cosmic variance.} 
\label{dgwd63}
\end{figure*}

The results for the distribution of dark gap widths 
for the two redshift ranges 
are plotted in Figure \ref{dgwd63}.
We find that the dark gap width distributions for 
ERM and LRM are essentially the same (accounting for the cosmic variance) 
for $W_{\alpha} < 40 $\ \AA, where $W_{\alpha}$ denotes the gap width in the 
Ly$\alpha$ forest.
However, one should note that 
for larger $W_{\alpha}$
the frequency of gaps differs substantially
for different models, with the difference being quite obvious
for the redshift range 6.0 -- 6.6.
As expected, LRM predicts an higher probability to find larger gaps
because of the neutral regions in the IGM. 
This difference in the two models at large gap widths can be used
as a possible discriminator between early and late reionization. However, it 
turns out that it
is possible to devise a more sensitive statistics for this purpose which 
we discuss in the next subsection.

\subsection{Distribution of largest gaps}  

\begin{figure*}
\psfig{figure=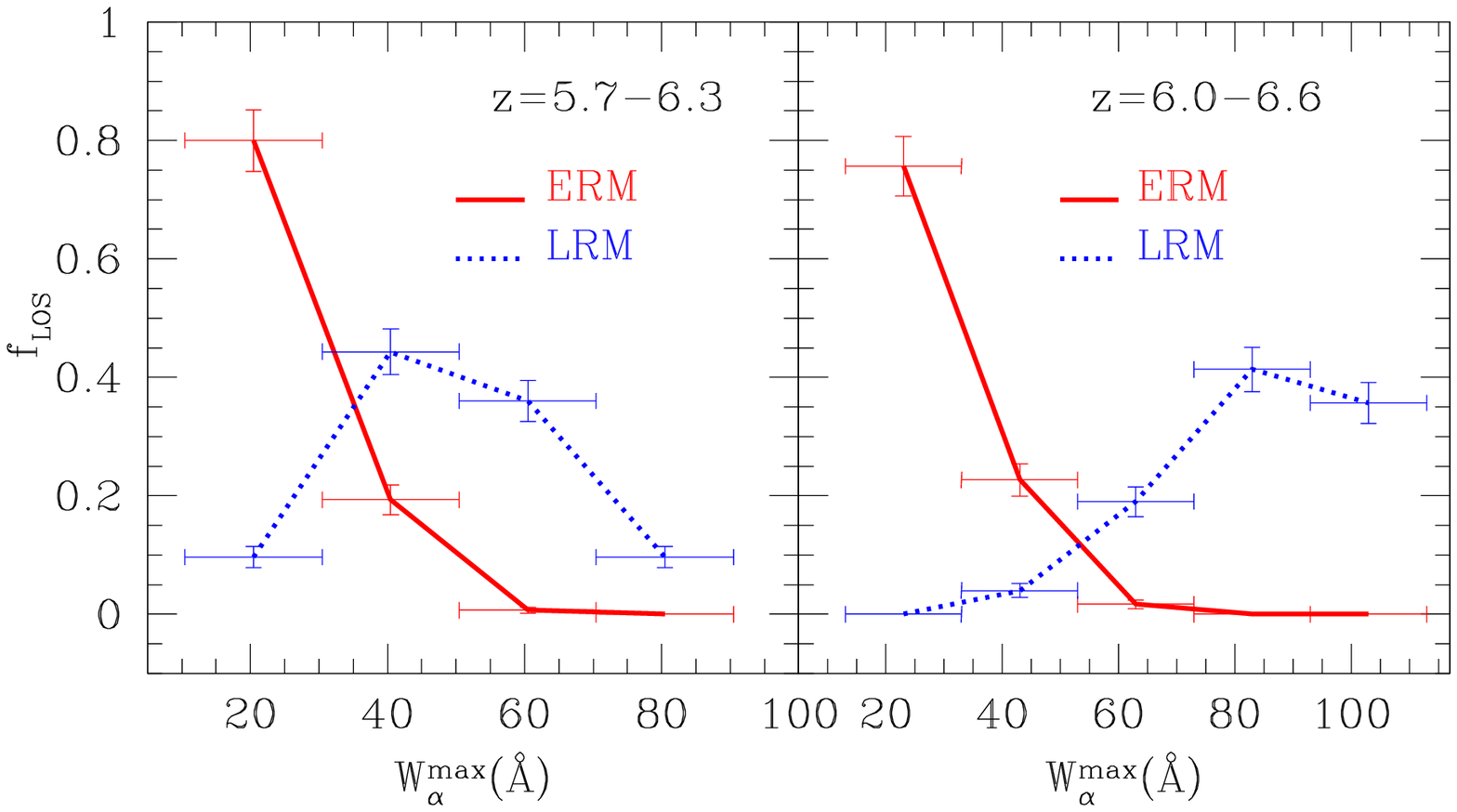,width=16.0cm,angle=0}
\caption{Distribution of the largest dark gap widths $W_{\alpha}^{\rm max}$
for 300 LOS in the redshift range 5.7 - 6.3 
(left panel) and 6.0 - 6.6 (right panel) for ERM (solid red line) and LRM 
(dotted blue line). The vertical error bars denote the cosmic variance;
the horizontal error bars show the bin size.} 
\label{pmax63}
\end{figure*} 

In this Section, we present the most sensitive diagnostic 
for distinguishing between different reionization scenarios 
using the Ly$\alpha$ forest. We calculate the width of the largest 
gap $W_{\alpha}^{\rm max}$
for each of the 300 LOS generated from our models, and then 
compute the fraction of LOS having a particular value of the largest 
gap width. It is clear from the discussion on Figure \ref{largest1}
that the typical size of the largest gap along a LOS
will be much larger in the LRM compared to ERM.
The 
fraction of LOS having a given value of largest gap is shown in Figure 
\ref{pmax63} with the corresponding cosmic variance.
It is clear that the distributions for the two models differ substantially; 
in particular, 
one should find gaps with $W_{\alpha}>50$ \AA \ for the 
LRM along $\sim 35$ per cent of the lines of sight, while if the universe 
is ionized early, there should be {\it no} 
line of sight with a dark gap
width $>$ 50 \AA. This is a very stringent result, and
can be used to rule out the early reionization
scenario from observational data. 
Similarly, the absence of dark gap widths $>$ 50 \AA \ can be used to rule 
out late reionization scenarios.

As expected, the difference between the two reionization models
is more drastic in the highest redshift range 6.0 -- 6.6.  In 
particular, we expect nearly half the lines of sight to have a gap
of width as large as 80 \AA \ if the universe is in the pre-overlap stage, 
while no such line of sight should be observed if the IGM is ionized.
Even if we take the statistical errors 
into account, we find that in order to validate 
the late reionization hypothesis, at least 40 per cent 
of the lines of sight should
have dark gaps larger than 70 \AA \ in the redshift range 6.0 -- 6.6.

At present SDSS has already observed 9 QSOs above redshift 6, and 
thus one should be able to compute the distribution 
of the largest gap widths. As an example, a visual inspection of the 
spectra of QSO SDSS J1030+0524 (which shows the 
darkest GP trough till date) reveals that the size of the 
largest dark gap is $< 40$ \AA, which can be well 
explained both by ERM and LRM. 

\subsection{Peak Width Distribution (PWD)}   

\begin{figure*}
\psfig{figure=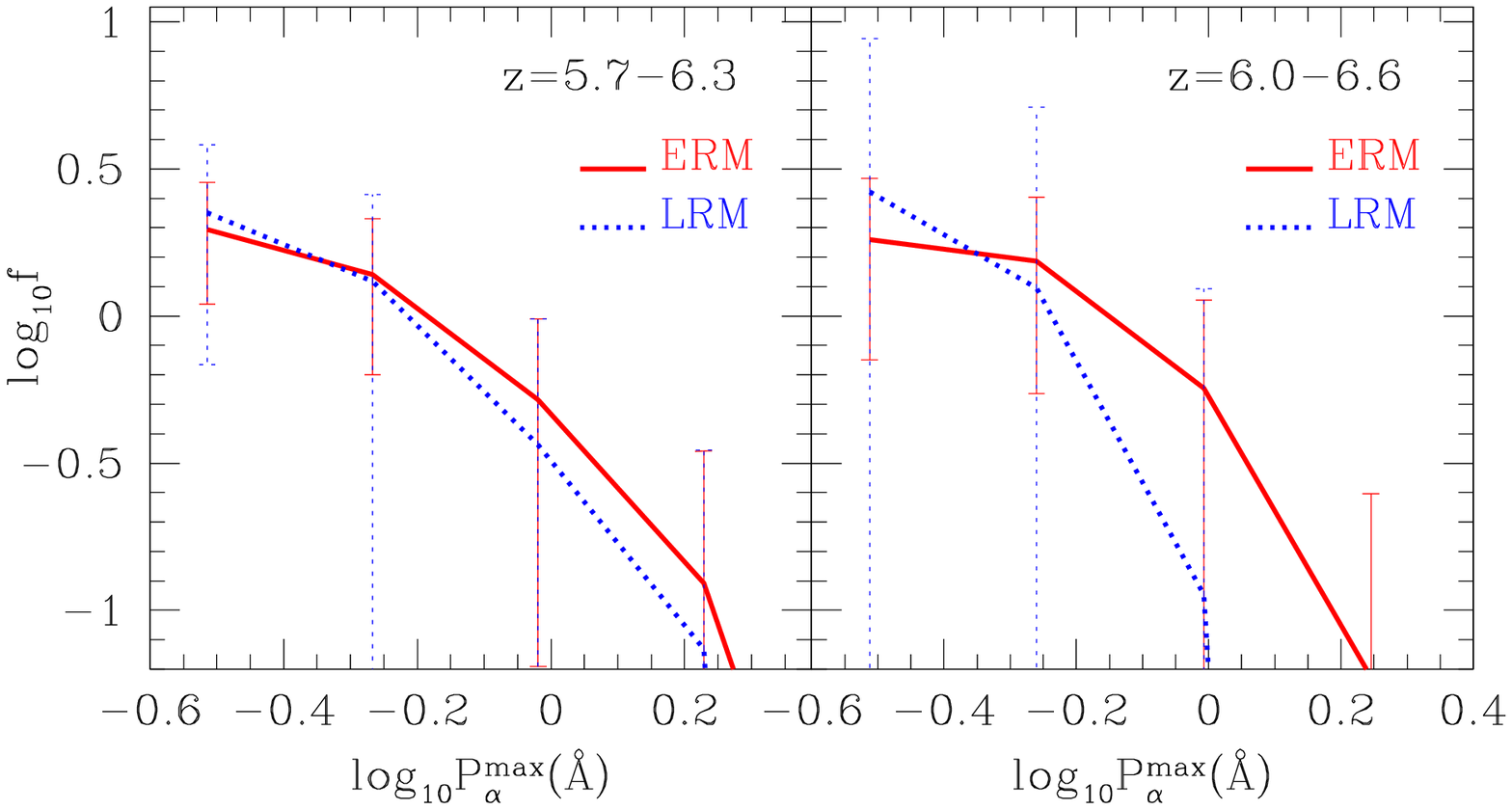,width=16.0cm,angle=0}
\caption{PWD at redshift ranges 5.7--6.3 (left panel) and 6.0--6.6 
(right panel). Solid (red)
and dotted (blue) lines represent the ERM and the LRM, respectively. 
For each model, the thick 
line is the average over 300 LOS, the error bars denote 
cosmic variance.} 
\label{pwd66}
\end{figure*} 

\begin{figure*}
\psfig{figure=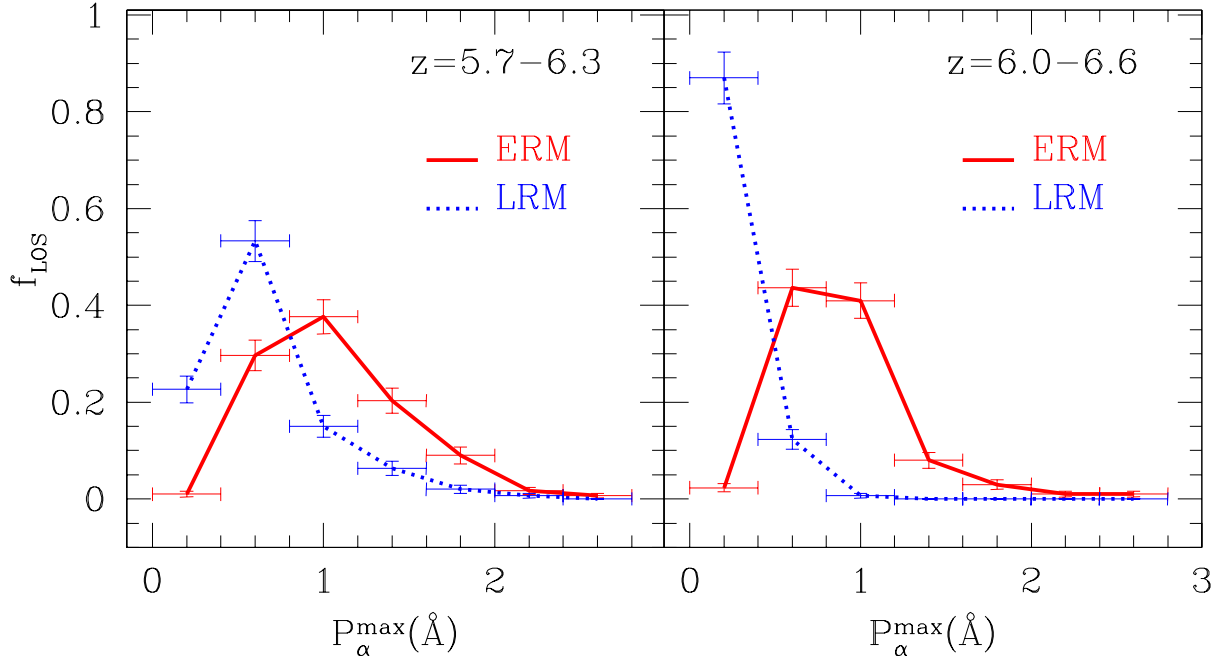,width=16.0cm,angle=0}
\caption{Same as in Figure \ref{pmax63} but for largest peak widths
$P_{\alpha}^{\rm max}$.} 
\label{peakmax63}
\end{figure*}

\begin{figure*}
\psfig{figure=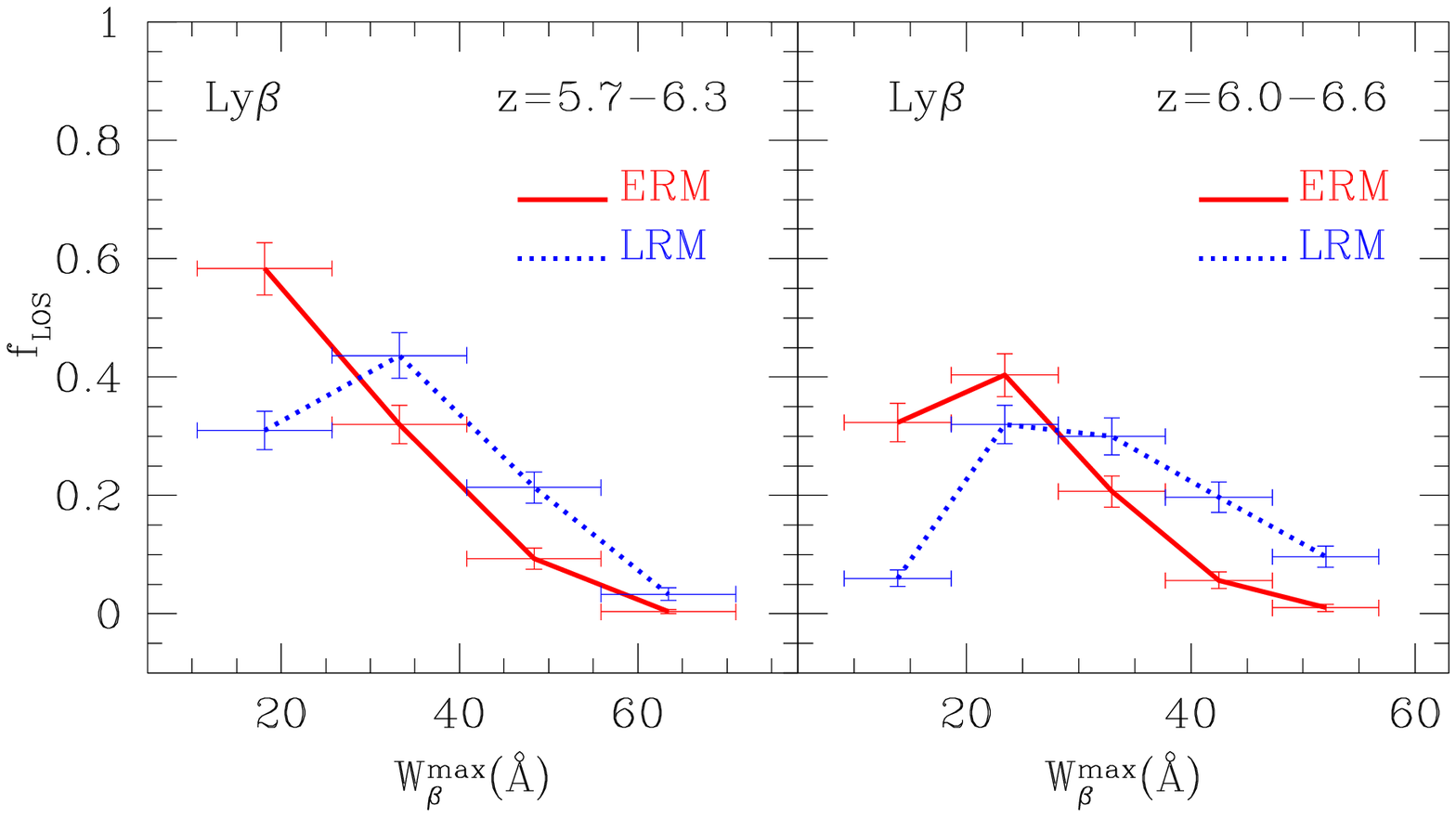,width=16.0cm,angle=0}
\caption{Same as in Figure \ref{pmax63} but for the Ly$\beta$ region.} 
\label{betagapmax}
\end{figure*}

Having identified a very useful statistics, the DGWD, we now introduce 
another possible analysis which can be thought as complementary 
to the gap statistics.
The Peak Width Distribution (PWD) allows us 
to measure the frequency and the width of those regions of the spectra 
characterized by a high transmission, i.e., a 
flux between 0.08 and 0.8  over rest frame 
wavelength intervals greater than 0.2 \AA \ (which is roughly equal to
6 pixels of our rebinned spectra). Visually these regions 
would appear as isolated spikes in the spectra at high redshifts 
(in this sense the terms ``spike'' and ``gap'' can thought of
as equivalent).
The lower threshold flux is same as the one used 
as the upper threshold 
in the DGWD analysis, while the upper limit value of 0.8 have been chosen 
in order to avoid regions of full transmission which could 
be affected by atmospheric absorption (though 
this effect does not seem to have much effect on the statistics).
Figure \ref{pwd66} shows the results for the PWD 
in the redshift ranges 5.7 -- 6.3 and 6.0 -- 6.6, respectively
with $P_{\alpha}$ denoting the width of the peak.
Similar to the DGWD, it is clear that for small peak widths, 
the errors are too large and thus do not allow to use this statistics 
for discriminating between the two models; however the distributions are quite 
different for peak widths larger than 1.2 \AA. This suggests that the 
fraction of lines of sight having a largest peak width
of a given value could be used as another discriminating statistics
between the two models. 

The results for the fraction of lines of sight
with a given value of the largest peak width $P_{\alpha}^{\rm max}$
are plotted in 
Figure \ref{peakmax63}.
It seems that at $z = 5.7 - 6.3$, the probability 
of finding a line of sight having a peak width larger than 1.2 \AA \  
is negligible in the LRM, while in an ERM 
peaks of this size seem to be present for 20 per cent of the lines of 
sight. The same effect is more evident in the redshift range $6.0-6.6$: there 
ERM predicts peaks of width $\sim 1$\AA \ in 40 per cent of the lines of sight;
on the contrary, the LRM predicts no peaks larger than 0.8 \AA.

We believe that
the distribution of peak widths can be used in a complementary way with
the dark gap width statistics for constraining the ionization state 
of the IGM at $z \gtrsim 6$.

\subsection{Results for the Ly$\beta$ region}

In addition to the Ly$\alpha$ forest, one can also use the
Ly$\beta$ region of the absorption spectra to constrain the 
ionization state of the IGM at high redshifts. The advantage 
of using the Ly$\beta$ absorption lines is that the absorption 
cross section is lower than the Ly$\alpha$ one, and hence one finds
some features of transmission within the spectra in Ly$\beta$ region 
even when  Ly$\alpha$ transmission is zero.
In this Section, we present our predictions for the 
Ly$\beta$ forest at $z > 6$.

We have calculated the DGWD for the Ly$\beta$ forest in the 
redshift ranges of interest and found it to be quite similar to the 
Ly$\alpha$ case. The distribution does show some differences 
between the reionization models at high values of gap widths, though
the difference is not as evident as in the case of Ly$\alpha$. 
We plot the fraction of LOS having a largest gap width 
of a given value in Figure \ref{betagapmax}, which corresponds to 
Figure \ref{pmax63} for Ly$\alpha$. Though the ERM and the LRM 
differ in their distributions for Ly$\beta$ regions, we find that 
it is not as discriminating as in the case of Ly$\alpha$ in the 
redshift intervals considered here. 

We have also computed the PWD distribution for the Ly$\beta$ region
of the spectra. The broad conclusions are similar to those obtained 
from Ly$\alpha$ regions, though the discrimination between LRM and ERM
is reduced in the case of Ly$\beta$.
However, the usefulness Ly$\beta$ 
statistics lies in the fact that these can be used as an independent check
for the reionization models.

\subsection{Dark gaps in both Ly$\alpha$ and Ly$\beta$ regions}

\begin{figure*}
\psfig{figure=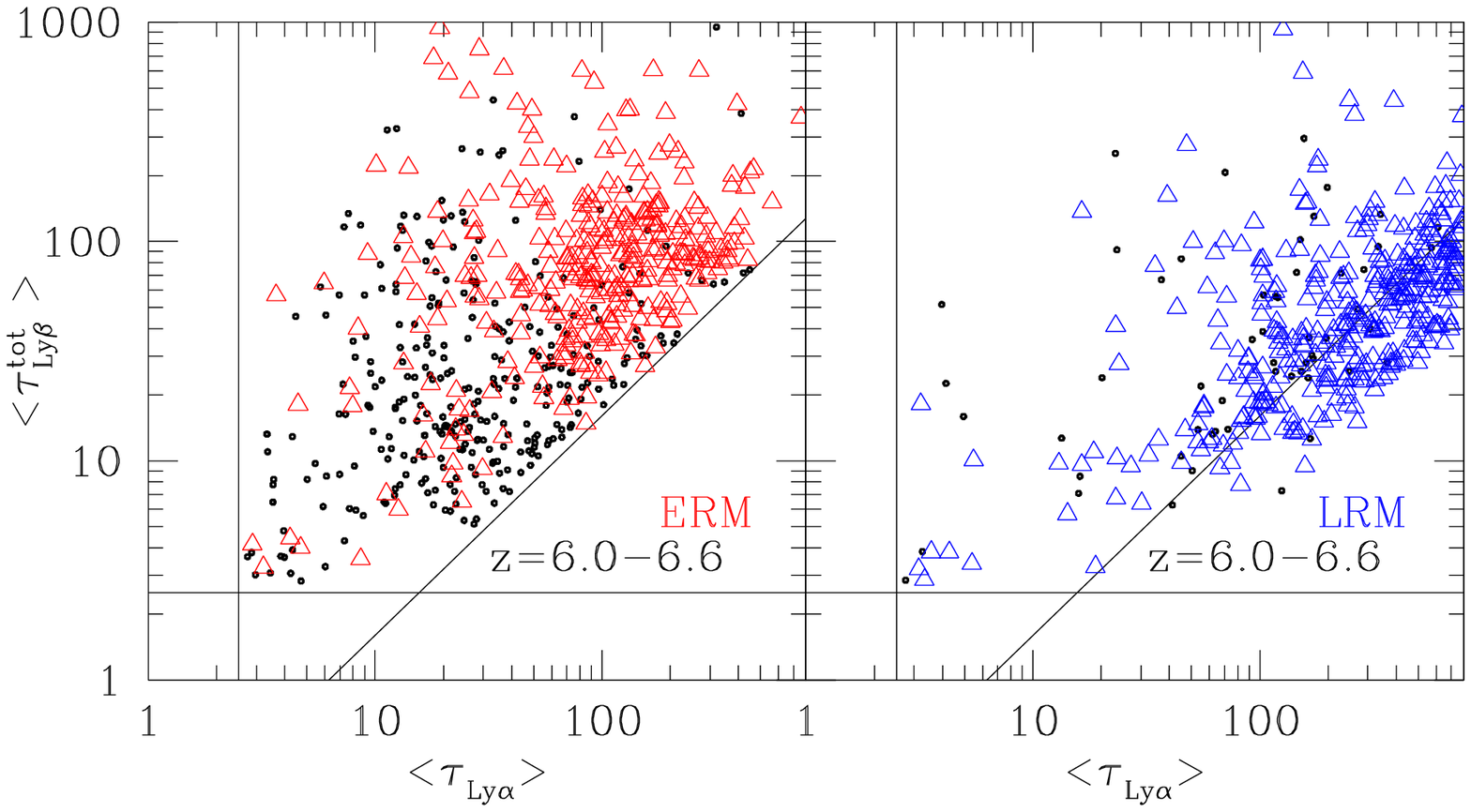,width=16.0cm,angle=0}
\caption{Scatter plot of the mean Ly$\alpha$ and the mean Ly$\beta$ 
optical depths for each dark gap. Points and triangles
represent dark gap 
such that 
$0\le \log_{10} [{\rm Max}\{W_{\alpha}, W_{\beta}\}] \le 0.5$ and 
$1.5\le \log_{10} [{\rm Max}\{W_{\alpha}, W_{\beta}\}] \le 2$, respectively, 
where ${\rm Max}\{W_{\alpha}, W_{\beta}\}$ is the width of the largest
between the Ly$\alpha$ and Ly$\beta$ dark gaps.
The solid lines parallel to the axes 
represent Ly$\alpha$
and Ly$\beta$ optical depths equal to 2.5, which acts as the lower
limit for defining gaps. The slope of the slanted
solid line is equal to the ratio $(f_{{\rm Ly}\beta}\lambda_{{\rm Ly}\beta})/
(f_{{\rm Ly}\alpha}\lambda_{{\rm Ly}\alpha})=0.16$.} 
\label{ska}
\end{figure*}

In this Section we study the presence of dark gaps in both the Ly$\alpha$ 
and the Ly$\beta$ regions of the absorption spectra. 
In Figure \ref{ska} we show the mean Ly$\alpha$ 
against 
the mean Ly$\beta$ optical depth for different dark gap lengths, for both 
reionization models. Points and triangles in the figure 
represent dark gaps such that 
$0\le \log_{10} [{\rm Max}\{W_{\alpha}, W_{\beta}\}] \le 0.5$ and 
$1.5\le \log_{10} [{\rm Max}\{W_{\alpha}, W_{\beta}\}] \le 2$, respectively, 
where ${\rm Max}\{W_{\alpha}, W_{\beta}\}$ is the width of the larger 
dark gap between the Ly$\alpha$ and Ly$\beta$.  It is obvious from the 
figure (and also stressed by \citeNP{pn05}) that 
larger dark gaps correspond to higher optical depths in both models.

There is one more interesting point to be noted from the figure.
Conventionally the total 
Ly$\beta$ optical depth is obtained from the Ly$\alpha$ one, using the 
following relation:
\begin{equation}
\tau^{\rm tot}_{{\rm Ly}\beta}(z)=0.16\tau_{{\rm Ly}\alpha}(z)
+\tau_{{\rm Ly}\alpha}(z_{\beta}),
\label{lybetano}
\end{equation}
which assumes that $\tau_{{\rm Ly}\beta}(z) = 0.16\tau_{{\rm Ly}\alpha}(z)$.
This assumption is true for low column density systems 
when the line profile of absorption is determined by the velocity
field and is same for Ly$\alpha$ and Ly$\beta$. In this case, the points 
in the $\tau^{\rm tot}_{{\rm Ly}\beta} - \tau_{{\rm Ly}\alpha}$ plane
will be strictly bound by a lower envelope, which will correspond to 
a straight line having a slope of 0.16. This bound is shown in the figure
as the slanted 
solid line. Note that for ERM there are truly no points below this line.
On the other hand, for LRM there are a lot of points below the 
solid line with slope 
0.16. This is related to the fact that the absorption from 
neutral regions present 
in the LRM cannot be described by a simple Gaussian profile and one
has to take into account the effect of damping wings.
This means that, as already discussed in Section \ref{voigt}, the usual 
adopted way to compute the total Ly$\beta$ 
optical depth from the Ly$\alpha$ one using equation (\ref{lybetano}) 
it is not appropriate in general, particularly
when the neutral fraction of the gas is high and
the Lorentzian part of the line profile 
becomes important.

\subsection{Variations in the Late Reionization Models}
\label{variations}

\begin{figure*}
\centerline{\psfig{figure=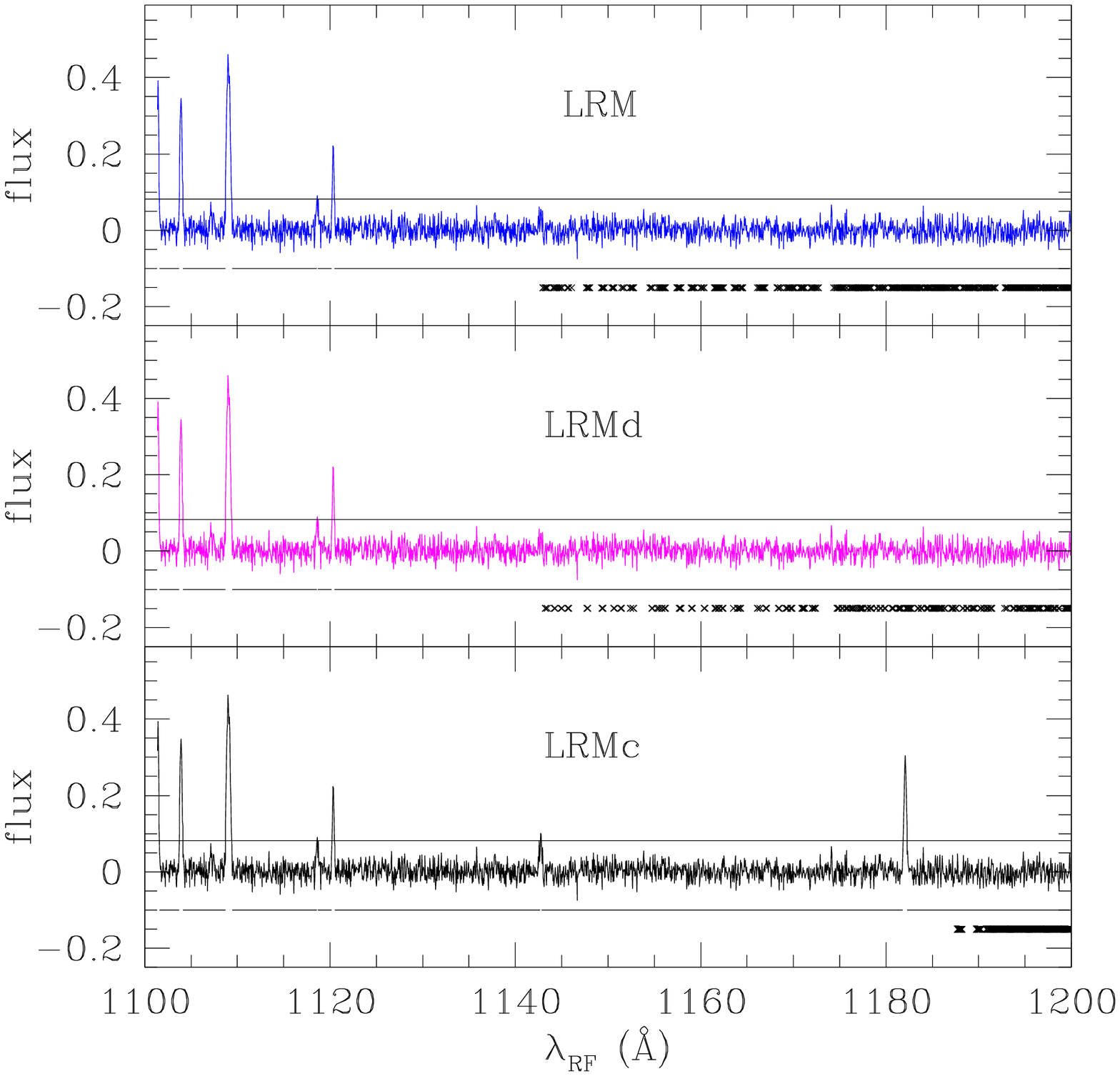,width=16.0cm,angle=0}}
\caption{Simulated 
spectra for three different models of late reionization. The top 
panel shows
a line of sight spectrum for LRM (same as in the 
bottom panel of Figure \ref{largest1}); the middle and bottom
panels show spectra along the same line of sight (i.e., having the same
density distribution) for LRMd and LRMc
respectively. The black lines 
plotted immediately below the spectra show the regions 
identified as gaps.
The positions of the neutral pixels are identified by crosses.}
\label{3spec}
\end{figure*} 

\begin{figure*}
\psfig{figure=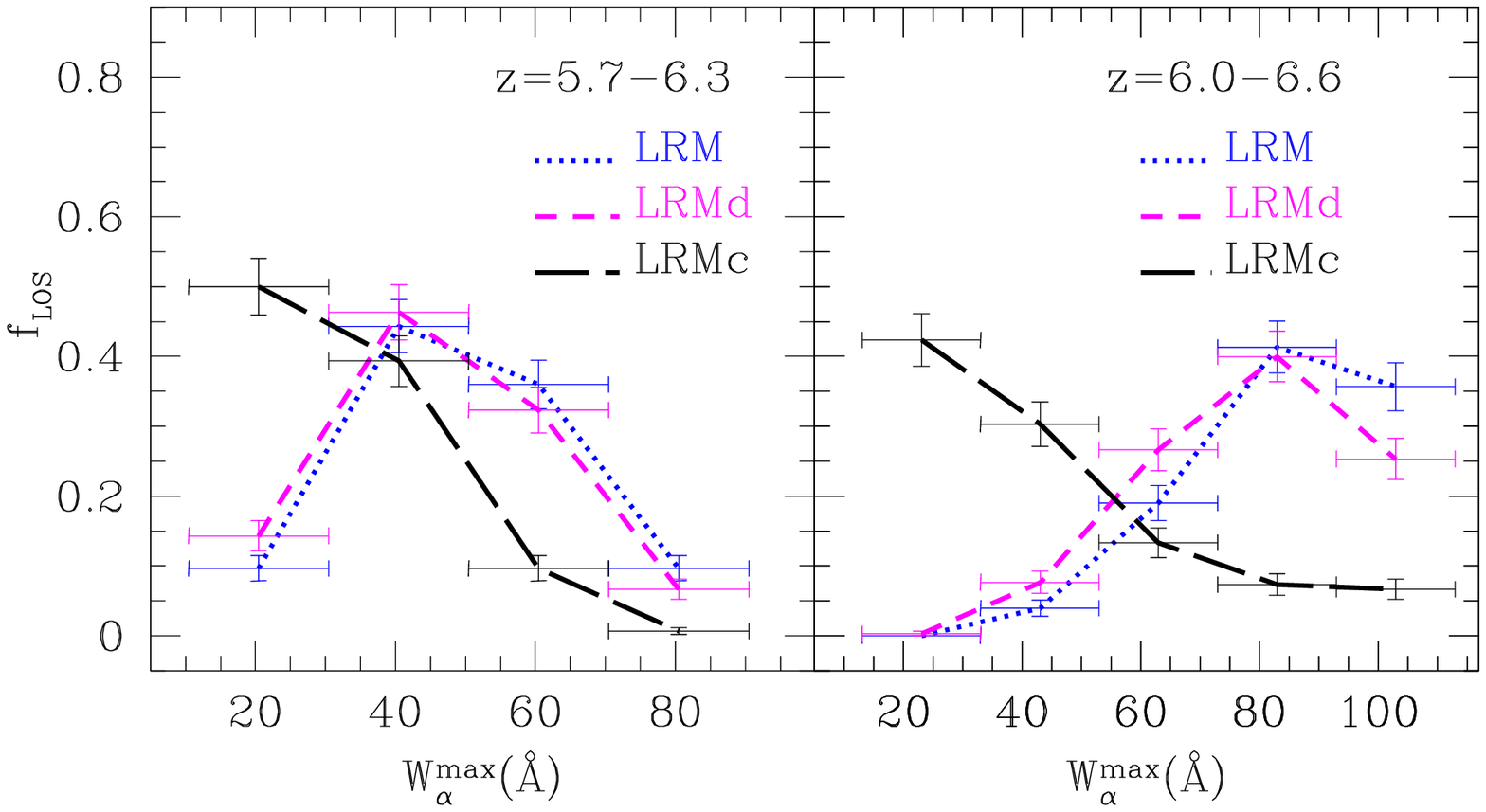,width=16.0cm,angle=0}
\caption{Same as in Figure \ref{pmax63} 
but considering only late reionization models. Dotted (blue), long-dashed 
(magenta) and
long-dashed (black)  
lines represent LRM, LRMd and LRMc respectively.} 
\label{pmax63_comp}
\end{figure*}

We now discuss certain other possibilities regarding the 
distribution of neutral regions along lines of sight. So far
we have been using LRM as the fiducial model for late reionization
which assumes that the neutral regions are 
distributed randomly, and they have {\it no} correlation
with the density field. However, this is not the
only possible way to distribute the neutral pixels. Hence we study two 
variations of the LRM which are named LRMd (d=density) and LRMc (c=clustered).
The LRMd is similar to LRM except that within a 
given redshift range the neutral pixels are correlated
with the density field with high density regions being preferentially neutral.
In the case of LRMc, we assume that the neutral regions are 
{\it maximally} clustered (i.e., they form a 
large coherent structure along the line of sight) 
when calculating the one-dimensional 
filling factor, $q_{\rm HII}$. This assumption represents the most extreme 
alternative to the LRM. In constructing our models we do not expect to recover 
exactly the real distribution of neutral regions; however, as we are 
considering the most extreme cases we expect the actual distribution to 
be somewhere between the two.
In the LRMc we find that the 
IGM is characterized by highly clustered large neutral regions 
(of lengths as large as few tens of comoving Mpc) and the 
correlation of these
regions with the density field does {\it not} have any 
effect on the simulated spectra. The technical details on how we generate these
models are discussed in Appendix \ref{q_hii}.

We start with the qualitative description of sample spectra for the
three models as shown in Figure \ref{3spec}. 
We recall that all the three panels have the same 
baryonic distribution and same value of $q_{\rm HII}$ 
along the line of sight (i.e., the 
number of neutral pixels, denoted by crosses, 
in the three panels are equal although it 
may not be visually obvious).

The first point to note is the similarity between the LRM and LRMd, implying
that the correlation of the neutral segments 
with the density field does not have much of an effect
on the Ly$\alpha$ forest. Although there are small differences in the 
actual positions of the neutral pixels in the two models, the 
gap widths are exactly similar for this line of sight. There are
differences in the widths of gaps for LRM and LRMd 
along other lines of sight but the variations are within the 
statistical errors. 

On the other hand, the spectrum for the 
LRMc is quite different from the spectra of LRM or LRMd. This is expected 
as the LRMc is very different from the other two in its physical
properties. As discussed in Appendix \ref{q_hii}, LRMc consists 
of very large neutral segments (up to 100 comoving Mpc), while the LRM and 
LRMd are characterized by numerous regions of smaller sizes.
Since the neutral regions are highly clustered in LRMc, they leave 
large volumes of ionized IGM. Consequentially we find that a high
fraction of lines of sight (about 70 per cent at $z = 5.7 - 6.3$ and
more than 80 per cent at $z = 6.0 - 6.6$) do {\it not} encounter
any neutral segments at all. 
Thus statistically LRMc is expected
to be the closest to ERM and should be quite different from LRM and LRMd.
Even when a line of sight encounters some neutral segments, it is more
likely that the segments are clustered at a few places forming large
regions of neutral IGM. This can be seen in the bottom panel
of Figure \ref{3spec} where we find that all the neutral pixels are
clustered at the highest redshift contrary to LRM and LRMd where 
the neutral pixels are distributed throughout the line of sight. 
This has a severe effect on the distribution of gaps and peaks. For example,
the peak around $\lambda_{\rm RF} = 1182$ \AA \ present in the ERM
(see top panel of Figure \ref{largest1}) is completely suppressed
in LRM and LRMd because of the randomly distributed neutral regions.
However, the peak is present in the LRMc because the neutral pixels are
distributed differently. This implies that LRMc would have gaps
of smaller widths compared to LRM and LRMd and thus would be closer to the 
ERM in its properties. Furthermore, it is clear from Figure \ref{3spec}
that 
{\it a large gap does not necessarily correspond to a large neutral region}. 
In fact 
we find that smaller regions of 
neutral hydrogen (of sizes $\lesssim 1$ comoving Mpc) dispersed along the line 
of sight are more effective in suppressing the flux and thus creating large
dark gaps in the
absorption spectra compared to the 
larger clustered regions. Probing such small regions
are quite difficult with cosmological simulations as they are close
to the resolution limits, thus semi-analytical studies can be of
more help in such cases.

We can carry out more quantitative comparisons between the different 
reionization models. We focus on the fraction of lines of sight
with a given value of largest gap width for the two redshift ranges.
Our results are plotted in Figure \ref{pmax63_comp}. As expected, 
the difference between LRM and LRMd is not statistically
significant. The fraction of larger dark gaps in both panels is
slightly lower in the LRMd with respect to the LRM. This result can be 
understood noting that in the LRMd the neutral regions are preferentially 
located in those pixels having an higher density, where the flux is already 
more likely suppressed. On the contrary the distribution for LRMc is quite 
different
and is more similar to ERM (see Figure \ref{pmax63}). The point
to note is that in spite of such extreme (and somewhat unphysical)
clustering, the LRMc is still different from ERM. For example, 10
per cent of the lines of sight have a largest gap width of 60 \AA \ 
(80 \AA) in the range $z = 5.7 - 6.3$ ($z = 6.0 - 6.6$) for LRMc which
are not present in the ERM. Thus even in the most extreme distribution
of neutral regions, the 
ionization state of the IGM can be determined using the dark gap statistics.

Moreover, it is also interesting to note that the distribution 
of the largest gap is quite different for LRM and LRMc -- thus
it might be possible to obtain some information on the clustering
of neutral regions. For example, 
in a quite realistic situation where we are provided with
only, say, 10 QSO spectra with emission redshift above 6, 
we expect to find in the LRMc
one LOS having a dark gap as large as 50 \AA \ (which would rule out ERM) 
and, in the same sample, 
at least 5
LOS whose largest dark gap does not exceed 30 \AA \ (which would rule out 
LRM). In this case, we could 
conclude at the same time that the universe is in the pre-overlap
stage and that the HI regions 
are highly clustered.

\subsection{Variations in the resolution and signal to noise ratio}

We have carried out extensive checks on our predictions by varying
different observational and instrumental artifacts. In particular,
we have varied the resolution and noise in the simulated spectra to 
verify if any of our conclusions change.

Our results presented in the previous sections are 
based on a resolution of 5300, which corresponds
to a FWHM of $\sim~60$~km~s$^{-1}$. 
We have checked our results for up to a resolution as high as 40000
(corresponding to a FWHM of $\sim 8$~km~s$^{-1}$), which 
is similar to what is expected in very high quality spectra. The results, 
particularly the gap and peak width statistics for the Ly$\alpha$ forest
do show some variations when the resolution is high. 
However, we find that none of our conclusions
get modified in a significant way.

For the noise, we have been using a Gaussian random variate having
variance $\sigma_{\rm noise} = 0.02$. Decreasing the value of 
$\sigma_{\rm noise}$ (which corresponds to higher signal-to-noise ratio)
has no effect on the gap and peak statistics. However, 
if we increase the value of $\sigma_{\rm noise}$ such that it 
becomes close to 
the flux threshold ($\sim 0.08$) used to define the dark gaps, we find that 
the occurrence of gaps (and peaks) 
changes drastically; there are various spurious spikes which arise 
because of high noise. Thus, to study
dark gaps in absorption spectra it is better to have a good signal-to-noise 
ratio; in case the signal-to-noise ratio is poor, it is necessary to 
use an higher flux threshold for defining the gaps.

\section{Conclusions}

In this work we have applied various statistical diagnostics to the 
transmitted flux of the Ly$\alpha$ (and Ly$\beta$) forest, with the aim 
of constraining cosmic reionization history. 
Two different reionization scenarios, based on self-consistent
models of \citeN{cf05}, have been considered: (i) an Early Reionization Model 
(ERM) characterized by a highly ionized IGM  at $z\lesssim 14$, and (ii) a 
Late Reionization Model (LRM) in which reionization occurs at $z\approx 6$.
These reionization histories are the result of different assumptions about the
type of ionizing sources considered. 
In both models, at redshifts $z < 6$ contributions to 
the UV background come from PopII 
stars and QSOs. The main difference between ERM and LRM 
is constituted by the presence of PopIII stars (not included in the LRM) 
which reionize the IGM at high redshift in ERM.
We note that for the ERM the electron scattering 
optical depth is $\tau_e=0.17$, in agreement 
with the WMAP value, whereas for the LRM $\tau_e=0.06$, a value more than  
2$\sigma$ away from the mean.
The main aim of this work is to quantitatively compare the predictions from
these two models, taken as representative of a wider class of
early or late reionization scenarios, with the highest quality observational 
data.  

First, we have extensively tested our results against available data at 
$z < 6$ 
and found that ERM and LRM are equally good at explaining the observational 
results. 
In particular, they reproduce very well the observed Gunn-Peterson optical 
depth evolution, 
the Probability Distribution Function of the transmitted flux, and 
the Dark Gap Width Distribution. This comparison allows us to draw a few 
conclusions:  
(i) the Ly$\alpha$ forest observations at $z<6$ are unable to discriminate 
early vs. late reionization scenarios;
(ii) the same data cannot exclude that reionization took place 
as early as by $z\approx 14$. 

In order to make progress higher redshift quasar spectra are necessary, 
which are likely to become soon available as SDSS is expected to   
find $\sim 20$ luminous quasars in the redshift range $6<z<6.6$. 
By extending our model predictions to higher redshifts we find that:
(i) The mean and the PDF of the transmitted flux are essentially useless 
to constrain the ionization state at $z \gtrsim 6$ as most of the
pixels are consistent with zero transmission (independent of the
ionization state), i.e. in practice these 
statistics probe the noise distribution;
(ii) the dark gap width distribution (DGWD) is very sensitive to the 
reionization history. We expect at least 30 per cent of the lines of sight 
(accounting for
statistical errors) in the range $z= 5.7-6.3$ to have dark gaps of widths 
$>50$ \AA \ (in the QSO rest frame) if the IGM is in the pre-overlap stage 
at $z \gtrsim 6$, while no lines of sight should have such large gaps if the 
IGM is already ionized.
The constraints become more stringent at higher redshifts. We find that
in order to discriminate between early and late reionization scenarios 
10 QSOs should be sufficient for the DGWD to give statistically robust results.
(iii) The statistics of the peaks in the spectra represents an useful 
complement to the dark gaps and can put additional constraints on the 
ionization state. 
As for the DGWD, we find that this statistics constrains reionization models 
more
efficiently at high redshifts. In particular, if the 
universe is highly ionized at $z\sim 6$, we expect to find peaks of width 
$\sim 1$\AA \ in 40 per cent of the lines of sight, in the redshift range 
$6.0-6.6$; on the contrary, the LRM predicts no peaks larger than 0.8 \AA.

As an independent check of the models, we have extended all the above 
statistics to Ly$\beta$ regions.
It turn out that this diagnostics is less powerful than the analog Ly$\alpha$ 
one to   
probe the ionization state of the IGM.
Moreover, since the Ly$\beta$ cross section is 5.27 times smaller than 
Ly$\alpha$ one, 
the flux is always higher in the Ly$\beta$ region than in the Ly$\alpha$ 
forest. 
This implies that to obtain Ly$\beta$ constraints as stringent as those from
Ly$\alpha$, requires the analysis of QSOs spectra for $z > 6.6$.

We would like to comment on some additional issues concerning LRM. 
As discussed in the text, the hydrogen distribution in the LRM for low density 
IGM is 
characterized by two distinct phases at $z\gtrsim 6$, namely an ionized and a 
neutral phase. To model this two-phase IGM we have studied 
different topologies of neutral 
regions. Interestingly, the main conclusions of our work remain unchanged
(see for instance Figure \ref{pmax63_comp}) irrespective of 
whether we assume that the positions of the neutral regions are completely 
random (LRM) or we correlate the HI regions along different lines of sight 
with the density field (LRMd). This result is basically due to the damping 
wings of neutral regions, which are able to suppress the flux in
 regions of the spectra that are fully ionized (See Figure
\ref{zoom}). 
On the other hand if the 
suppression of the flux does not necessarily correspond to the presence of 
neutral regions, it implies that QSO spectra might not be very useful 
to study in details the topology
of the neutral hydrogen. 

However it is still possible to get some
idea about the clustering of the neutral regions
{\it provided we know the evolution of the volume filling factor
of ionized regions reasonably well}.
We have studied an alternative distribution of the neutral regions, called
LRMc, where we assume that neutral regions form the
largest possible coherent structure along the line of sight (sometimes 
as large as 100 Mpc comoving which corresponds to 
almost $1/3$ of the box).
Because of such high clustering, large volumes of IGM are left ionized, 
resulting in a large fraction of lines of sight which do {\it not} encounter
any neutral region at all.
Consequently, the distribution of the largest dark gap widths is biased 
towards lower widths compared to LRM.
This means that the
statistics of the largest dark gaps could also 
give an idea of the clustering in the
HI regions.
Moreover, as is well known, the 21 cm signal from neutral
hydrogen is sensitive to distribution of
the HII regions \cite{fhz04,fzh04}. 
Hence 21 cm maps could be 
promising to study the correlation between neutral 
regions and to obtain
a more detailed and quantitative analysis of the size 
of neutral regions.
 
Comparing the LRM and LRMc, we also find that 
{\it a large gap does not necessarily correspond to a large neutral region}.
In fact smaller regions of 
neutral hydrogen (of sizes $\lesssim 1$ comoving Mpc) dispersed along the line 
of sight are more effective in suppressing the flux (because of damping wings) 
and thus creating large dark gaps in the absorption spectra compared to the 
larger clustered regions. Probing such small regions
is quite difficult with cosmological simulations as they are close
to the resolution limits, thus semi-analytical studies can be 
more helpful in such cases.
Our method, in fact,  does not suffer of spurious resolution effects. 
At high redshift, the 
length of dark gaps can be $\gtrsim 60$ Mpc and hence the analysis requires a 
large sample of very long lines of sight.
In order to create realizations of such long lines of sight, numerical
simulations typically sample different regions of the box more than
once (the so-called ``oversampling'' effect; \citeNP{pn05}) or combine 
various spectra of smaller sizes end-to-end (F02). It is 
difficult to obtain the distribution of very large gaps 
(which are much larger than the box sizes) 
from such procedures as multiple ray passages through the same box 
could produce spectacular spurious artifacts in the gap
statistics. For example, we find a much better
match with the observations of dark gap width 
distribution when compared to the simulations of \citeN{pn05}, who have
used a box of size 6.8 $h^{-1}$Mpc.

However, our method suffers from some limitations which are 
worth noting. First, we are not able to tackle the non-linearities
in any self-consistent formalism -- instead we assume a density
distribution for the baryons (lognormal, in this case). Since the 
Ly$\alpha$ and Ly$\beta$ forests in the QSO absorption spectra arise from 
mostly quasi-linear regime, the approximation should be reasonable
for computing the transmitted flux.
Second, it is nearly impossible to include full radiative transfer effects 
in our
computation of the distribution of the neutral regions and
also we are not 
able to take into account the clustering 
of sources which is crucial to understand the properties of ionized bubbles.
However it is most likely that the location of ionizing sources 
might not be significantly
correlated with neutral regions, particularly when 
one is dealing with high values 
of filling factor, as in our case (see, for example, the maps in 
Figure 1 of \citeNP{cfw03}).  
Anyway it would be interesting to combine our approach in the distribution of 
neutral regions with radiative transfer simulations for a more detailed 
analysis of the absorption spectra, particularly in the vicinity of the QSO 
(which corresponds to a highly non-linear structure which is beyond the 
validity of the lognormal approximation).

In the future, it would be most interesting to check our predictions 
against a large sample of high signal-to-noise 
QSO data at redshift $>6$. Our results 
show that in that case the dark gap statistics would provide a robust and 
independent probe of the reionization history.

\section*{Acknowledgment}
We acknowledge useful discussions with G. Becker, L. Cowie, S. Cristiani,
X. Fan, M. Haenhelt, A. Nusser, W. Sargent, J. Schaye, A. Songaila, M.
Strauss, T. Theuns, M. Viel.  A special thank goes to N. Gnedin for
generously providing us with the HydroPM code. We thank the referee
for useful comments which have helped in improving the clarity
of the paper.

\bibliography{}
\bibliographystyle{mnras}

\appendix

\section{Lognormal approximation vs. simulations}
\label{lognormal}

\begin{figure*}
\centerline{\rotatebox{270}{\psfig{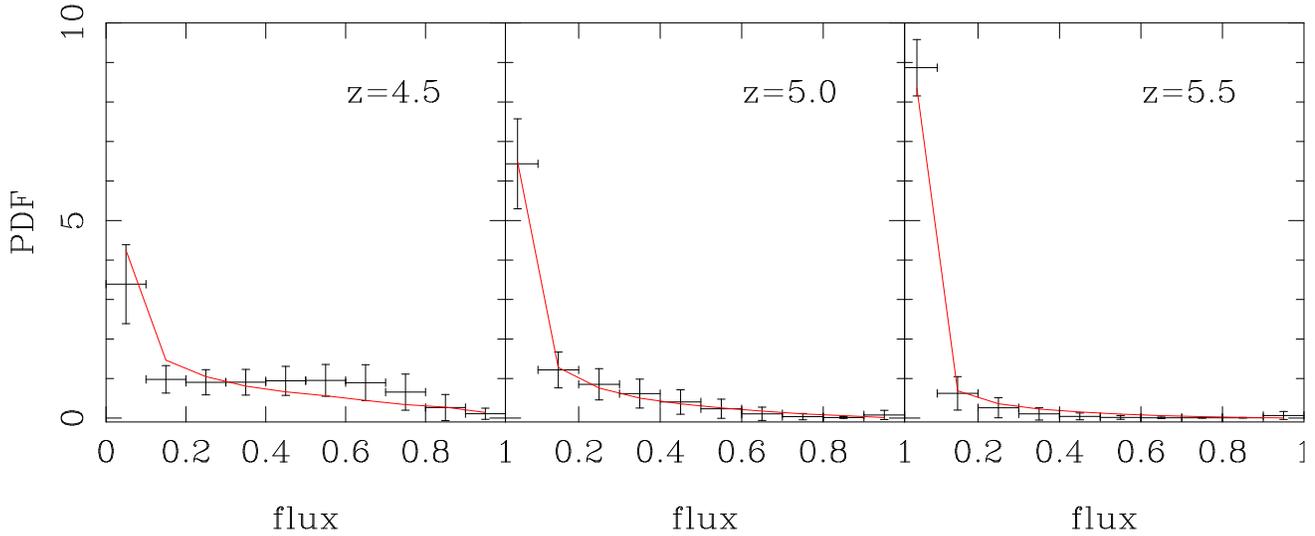}}}
\caption{Comparison of the PDF of transmitted flux 
obtained from lognormal model, shown as solid lines, with that obtained from 
HydroPM
simulations, shown as points with error bars. The vertical error bars 
represent dispersion along different lines of sight, while the horizontal 
error bars denote the bin size.
We show the results for three redshifts
$z = 4.5$ (left panel), $z = 5.0$ (middle panel) and $z = 5.5$ (right panel).} 
\label{lnhydropdf}
\end{figure*} 

\begin{figure*}
\centerline{\rotatebox{270}{\psfig{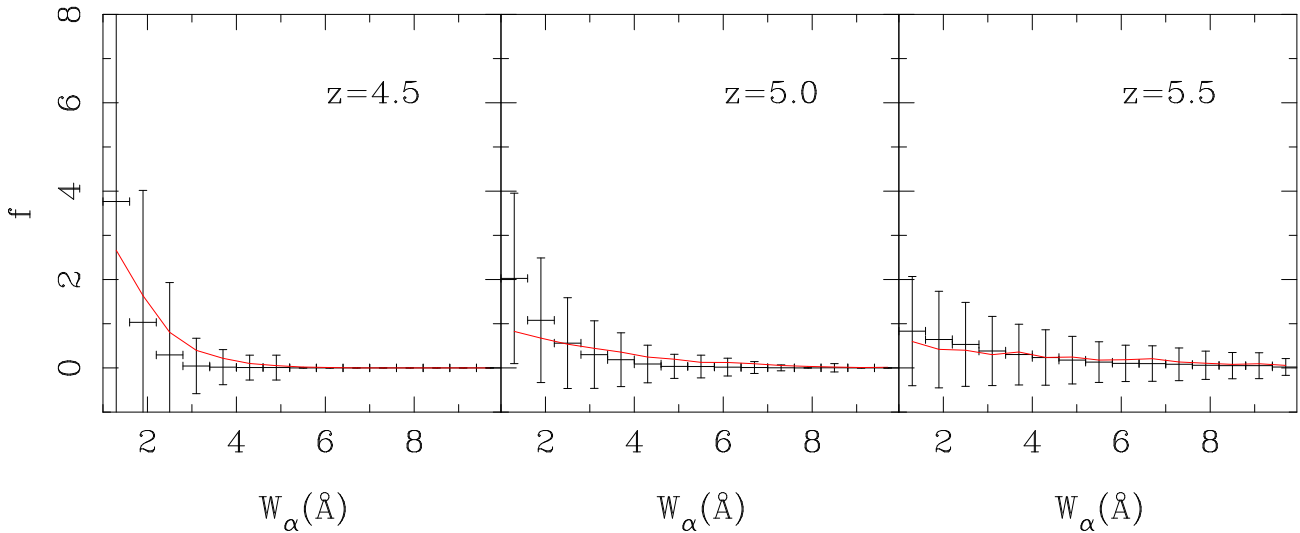}}}
\caption{Same as in Figure \ref{lnhydropdf} but for DGWD.} 
\label{lnhydrogap}
\end{figure*} 

\begin{figure*}
\psfig{figure=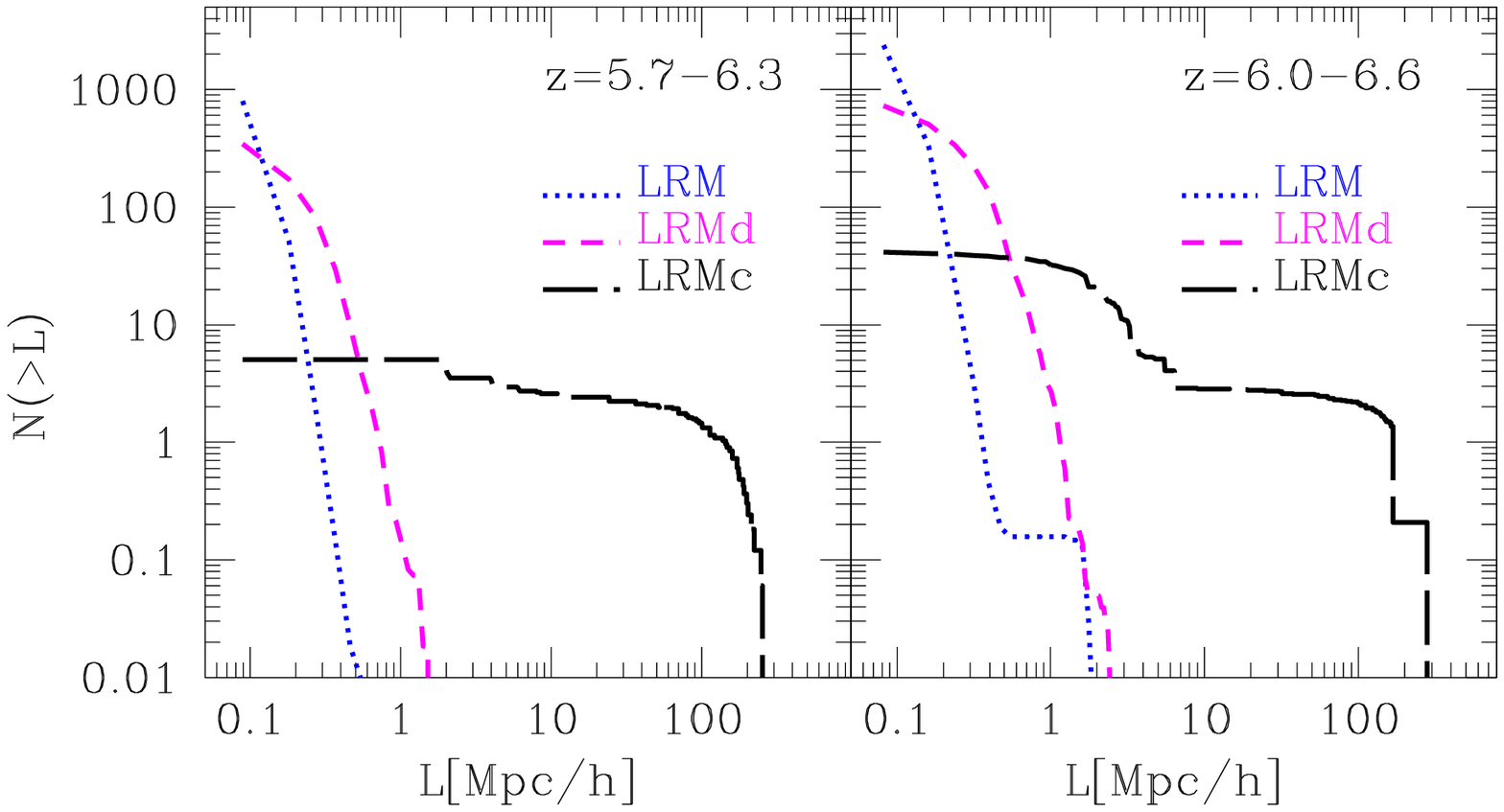,width=16.0cm,angle=0}
\caption{Cumulative distribution of the lengths 
of the neutral regions in the redshift ranges 5.7--6.3 (left panel) 
and 6.0--6.6 (right panel) for three different reionization models
as indicated in the figure.} 
\label{frapix}
\end{figure*}  

In this Appendix we compare the various Ly$\alpha$ 
flux statistics, namely, the PDF
and DGWD, computed using the lognormal model with those
obtained from numerical simulations. Of course, a 
thorough verification of the lognormal approximation would require a
comparison with full hydrodynamical simulations; 
however, since it has been found that the HydroPM 
simulations are able to reproduce most of the physical
properties of the Ly$\alpha$ forest and are computationally 
much less expensive, we shall 
restrict ourselves to HydroPM 
simulations for the purpose of this work. We should mention here 
that the comparison presented here is mainly to justify the 
lognormal approximation for doing statistics with Ly$\alpha$ forest 
at $z \approx 6$ -- this is not intended to be a rigorous
justification for the lognormal approximation for the baryonic density
field.

While comparing our model with HydroPM 
simulations, one should keep in mind that the two models 
are \emph{not at all} similar in their treatment of the 
reionization histories. While our models treat the reionization 
as an extended and gradual process with different ionized regions
overlapping gradually, the simulations consider an abrupt
reionization. 
Since sorting out such issues require 
much detailed effort, we 
restrict the comparisons in the post-reionization epoch 
(i.e., $z < 6$) and  
choose the parameters (cosmological model, 
$T_0, \gamma, \Gamma_{\rm HI}$) in our code 
in a manner that they have the 
same values as HydroPM simulations at $z < 6$. This allows us to have a fair
comparison between the lognormal model and numerical simulations
with uncertainties due to different reionization histories
under control. Note that, for
the comparison in this Appendix, we have {\it not} 
used the reionization models (LRM or ERM) described
in the paper -- in fact, we have used the values from the HydroPM code.

We have run HydroPM 
simulations with $128^3$ particles in a $12.8 h^{-1}$ Mpc box
having a mesh size of $100 h^{-1}$ kpc. At various 
redshift outputs, random LOS were chosen along different
directions and the corresponding density and velocity
fields were calculated. It is then straightforward 
to obtain the transmitted flux along the LOS given the values of 
different parameters related to the IGM. In the case of the lognormal model, 
we use the same procedure outlined in Section 2. We use 
a box size and resolution similar to what is used in the 
numerical simulations so that the resolution effects are 
not substantial. We have \emph{not} added any observational artifacts
(smoothing, noise etc) so that the comparison is restricted
to physical properties of the IGM. We compute the PDF and DGWD 
in a manner similar to what is described in the main text. 

The results for PDF and DGWD are shown in Figures \ref{lnhydropdf} and 
\ref{lnhydrogap} respectively
for three redshifts 4.5, 5.0 and 5.5. At redshifts lower than 4.5, the 
frequency of gaps reduces considerably and hence, in some sense, the 
usefulness of the gap statistics becomes irrelevant. On the other hand, at 
higher redshifts, the size of the gaps become of the order or larger
than the box size of the HydroPM simulations, and hence one needs to 
use considerably larger box sizes to carry out the comparison. Furthermore, 
at redshifts closer to 6, the physical properties of the IGM might be 
affected by details of the reionization history (particularly if the 
reionization is late), and hence we restrict ourselves to $z < 5.5$.

As can be seen from the Figure, the agreement between lognormal 
approximation and HydroPM simulations is excellent at $z = 5.5$, 
both for the PDF and the DGWD. The agreement is slightly less 
for $z = 5.0$ and is acceptable (within 1$\sigma$) at $z = 4.5$. 
It is thus clear that the lognormal model can be used reliably for 
Ly$\alpha$ transmitted flux statistics, particularly at high redshifts.

\section{Volume filling factor of ionized regions}
\label{q_hii}

In Sections 2.1 and 2.2 physical properties of different reionization models
have been discussed. In particular it results that in the Late Reionization 
Model (LRM) the IGM is characterized by two distinct phases at $z\gtrsim 6$,
namely an ionized and a neutral phase.
The aim of this Appendix is to explain how we distribute
the fully neutral regions along different lines of sight in the 
LRM, taking into account
the scatter and evolution in the 
volume filling factor $Q_{\rm HII}(z)$ of ionized regions. 
The method consists of two parts which are described in the 
two following subsections:

\subsection{Calculation of the one dimensional filling factor $q_{\rm HII}(z)$}

This involves the geometrical translation of 
the three-dimensional filling factor 
$Q_{\rm HII}(z)$ to a distribution of the 
one dimensional filling factor $q_{\rm HII}(z)$
along different lines of sight. This calculation can be performed once 
$Q_{\rm HII}(z)$ and the geometry of the neutral regions are known.

We start the procedure with a three-dimensional box 
with two spatial directions (representing two directions on the sky) 
and one direction along the redshift axis (representing the direction 
along the line of sight).
The spatial extent of the box can be arbitrary as we are 
only doing a geometrical exercise.
We divide the box into (thin) redshift slices and within each 
slice distribute a number of neutral regions according 
to the value of $Q_{\rm HII}(z)$ at that redshift. Note that 
this procedure automatically takes into account the 
evolution in the volume filling factor. However, the value of 
$Q_{\rm HII}(z)$ at a redshift does not contain information 
of the typical sizes and shapes of the neutral regions.
We thus consider the two most extreme cases, one in which the 
neutral regions are distributed in a completely random manner
with no clustering, while 
in the other case we put {\it maximally correlated} 
neutral regions which are of the largest
size allowed by the value of $Q_{\rm HII}(z)$ at that given redshift. 
In the first case, the 
box is characterized by numerous neutral regions of very small sizes, while
in the second one, the box consists of bigger neutral blobs 
representing the maximum clustering of neutral regions.
While in reality, none of these cases may represent the geometry
of the neutral regions, they nevertheless represent the two extremes 
and thus we are sure that the actual case is somewhere between these.

Once we have distributed the neutral regions within the 
three-dimensional box, we shoot numerous lines of sight through it. 
Each line of sight intersects different neutral regions at different 
redshifts, and thus we can calculate the one-dimensional
filling factor $q_{\rm HII}(z)$ along each of them. Thus given 
a single $Q_{\rm HII}(z)$, we build up a distribution 
of $q_{\rm HII}(z)$ characterizing each line of sight. 
Using this distribution, we 
are not only able to take into account the evolution in $Q_{\rm HII}(z)$, 
but also the intrinsic scatter (or, cosmic variance) 
in the distribution of neutral regions.

\subsection{Correlation between neutral regions and the density field}

Now that we have found out (at least) two ways of calculating 
$q_{\rm HII}(z)$ along a given line of sight, 
we have to accordingly distribute the neutral pixels along the same. For this,
it is essential that we know
whether the neutral regions have any correlation with the density
field. The Ly$\alpha$ forest arises mostly from baryonic overdensities
of a few ($\lesssim 5$) and it is not clear whether the neutral regions 
have any correlation with densities within such ranges (the correlation
is much more established in case of higher densities).
Hence we have tried both 
the options; in the first case, we have distributed the neutral pixels
along the line of sight without any consideration for the density 
field, while in the second case we distribute the neutral pixels such that
high density regions are preferentially neutral at the same time
preserving the evolution trend of $q_{\rm HII}(z)$. 

\subsection{Different LRMs}

Since we have two ways of calculating the one-dimensional filling
factor and furthermore have the freedom in choosing whether the  
neutral regions are correlated with the density
field, we can devise various LRMs which will cover 
all the extreme possibilities
of distributing the neutral regions.
\begin{itemize}
\item {\bf LRM:} While computing the distribution of the 
one-dimensional filling factor, we assume that the neutral regions are 
distributed randomly, and 
while distributing the neutral pixels along lines of 
sight, we assume that they have {\it no} correlation
with the density field. This acts as the fiducial model for our paper.

\item {\bf LRMd:} This is similar to LRM except that within a 
given redshift slice the neutral pixels are correlated
with the density field with high density regions being preferentially neutral.

\item {\bf LRMc:} In this model we want to reach the maximum level of
clustering for neutral regions. To do this we assume that neutral regions are
coherent structures (resembling filaments extended along the lines of sight) 
which preserve the evolution of the volume
filling factor. 
Once we assume this, we find that the 
IGM is characterized by highly clustered large neutral regions 
(of lengths as large as few tens of comoving Mpcs) and the 
correlation of these
regions with the density field does {\it not} have any 
effect on the simulated spectra.

\end{itemize}

To understand the physical properties of the different LRMs, it is useful to 
compute the distribution of sizes of the neutral regions
along all the LOS. The number of regions $N (> L)$ along different 
lines of sight having a length 
greater than $L$ for two redshift intervals of interest 
is plotted in Figure \ref{frapix}.
We have normalized the distribution such that the total 
length of the line of sight is $1 h^{-1}$ Gpc, so as to 
compare our results to those obtained from simulations
\cite{nbsl02}.

The first obvious fact to note by comparing the two panels of the 
Figure is that, for a given model, the number of neutral regions 
with comparatively larger sizes increases as one goes to higher redshifts. 
Furthermore, it is also clear that the LRMc consists of very large regions
($\sim$ 10--100 $h^{-1}$ comoving Mpc) as it represents the model with maximum 
clustering
of neutral regions. Among the other two models (LRM and LRMd), the LRMd 
has regions of relatively larger lengths; the reason is because of the 
correlation with density field. The LRM contains {\it no} clustering 
of the neutral regions and thus, as expected, is characterized by 
numerous regions of small sizes ($\lesssim$ few comoving Mpcs).

As an independent check on our procedure, Figure \ref{frapix} can 
be compared with Fig. 1 of \citeN{nbsl02} in a qualitative manner. 
\citeN{nbsl02} use N-body simulations coupled to a semi-analytical
galaxy formation model and various models of propagation of 
ionization fronts to calculate the sizes of neutral segments along
lines of sight. 
Since our method for calculating the reionization
history is quite different from theirs, it is not possible to carry out a more 
quantitative comparison. Even though our modelling of neutral regions 
predicts a 
larger number of small regions than simulations, we have to keep in mind that 
our resolution in computing the QSO spectra is 
much higher than simulations (in our case it is 
$\sim 0.04 h^{-1}$ Mpc while in the simulations of 
\citeNP{nbsl02} it is $\sim 0.55 h^{-1}$ Mpc). 
As also claimed by \citeN{nbsl02}, the lengths of 
neutral regions depend on the resolution adopted in the computation, i.e., 
increasing the resolution increases the number of small regions. As
a check, when we 
compute the distribution of neutral regions using a resolution close 
to \citeN{nbsl02} one, we find that our results are in good agreement with 
simulations in the LRM/LRMd cases,  thus supporting the
reliability of our approximate method of distributing neutral regions, while 
LRMc results to be quite distant from 
the simulation results as it 
produces segments as large as 100 $h^{-1}$ comoving Mpcs which are never seen
in simulations; nevertheless we study it as an extreme case.

We end this Appendix pointing to the fact that the numerical 
simulations cannot probe lengths much smaller than 
1 $h^{-1}$ comoving Mpc because of resolution effects; however 
with our
semi-analytical models, it is possible to probe neutral regions 
with sizes as small as 0.04 $h^{-1}$ comoving Mpc. Interestingly, it
turns out that these small regions, distributed randomly, are as 
effective as larger regions in suppressing the flux of the 
quasar spectra. This issue has been discussed in great detail in 
Section \ref{variations}.

\end{document}